\documentclass[sigconf]{acmart}
    
\usepackage{balance}
\usepackage{multirow}
\usepackage{subfigure}
\usepackage{color}
\usepackage{url}
\usepackage{amssymb}
\usepackage{pifont}
\usepackage{float}
\usepackage{amsmath}
\usepackage[ruled,linesnumbered]{algorithm2e}
\usepackage{algpseudocode}
\SetArgSty{textup}
\usepackage{pifont}
\usepackage{graphicx}
\usepackage{makecell}
\usepackage{balance}
\usepackage{array}

\makeatletter
\renewcommand*\env@matrix[1][*\c@MaxMatrixCols c]{%
  \hskip -\arraycolsep
  \let\@ifnextchar\new@ifnextchar
  \array{#1}}

\renewenvironment{bmatrix}
{{\ifnum`}=0 \fi\left[\env@matrix}
{\endmatrix\right]\ifnum`{=0 \fi}}
\makeatother

\newcolumntype{?}[1]{!{\vrule width #1}}

\AtBeginDocument{%
  \providecommand\BibTeX{{%
    \normalfont B\kern-0.5em{\scshape i\kern-0.25em b}\kern-0.8em\TeX}}}

\copyrightyear{2019}
\acmYear{2019}
\setcopyright{acmcopyright}
\acmConference[CIKM '19] {The 28th ACM International Conference on Information and Knowledge Management}{November 3--7, 2019}{Beijing, China}
\acmBooktitle{The 28th ACM International Conference on Information and Knowledge Management (CIKM'19), November 3--7, 2019, Beijing, China}
\acmPrice{15.00}
\acmDOI{10.1145/3357384.3358009}
\acmISBN{978-1-4503-6976-3/19/11}

\begin{document}

\title{Balance in Signed Bipartite Networks}

\author{Tyler Derr}
\affiliation{
  \institution{Data Science and Engineering Lab \\ Michigan State University}
}
\email{derrtyle@msu.edu}

\author{Cassidy Johnson}
\affiliation{
  \institution{University of the Pacific}
}
\email{c_johnson49@u.pacific.edu}

\author{Yi Chang}
\affiliation{
  \institution{School of Artificial Intelligence \\ Jilin University, China}
}
\email{yichang@jlu.edu.cn}

\author{Jiliang Tang}
\affiliation{
  \institution{Data Science and Engineering Lab \\ Michigan State University}
}
\email{tangjili@msu.edu}


\begin{abstract}
A large portion of today's big data can be represented as networks. However, not all networks are the same, and in fact, for many that have additional complexities to their structure, traditional general network analysis methods are no longer applicable. For example, signed networks contain both positive and negative links, and thus dedicated theories and algorithms have been developed. However, previous work mainly focuses on the unipartite setting where signed links connect any pair of nodes. Signed bipartite networks on the one hand, are commonly found, but have primarily been overlooked. Their complexities of having two node types where signed links can only form across the two sets introduce challenges that prevent most existing literature on unipartite signed and unsigned bipartite networks from being applied. On the other hand, balance theory, a key signed social theory, has been generally defined for cycles of any length and is being used in the form of triangles for numerous unipartite signed network tasks. However, in bipartite networks there are no triangles and furthermore there exist two types of nodes. Therefore, in this work, we conduct the first comprehensive analysis and validation of balance theory using the smallest cycle in signed bipartite networks - signed butterflies (i.e., cycles of length 4 containing the two node types). Then, to investigate the applicability of balance theory aiding signed bipartite network tasks, we develop multiple sign prediction methods that utilize balance theory in the form of signed butterflies. Our sign prediction experiment on three real-world signed bipartite networks demonstrates the effectiveness of using these signed butterflies for not only sign prediction, but paves the way for improvements in other signed bipartite network analysis tasks.
\end{abstract}


\maketitle

\section{Introduction}

Much of the data being produced today can be represented in networks. However, many networks exist that fall outside the range of being able to apply the general network analysis methods to them due to their added complexities in structure. One such type that has become increasingly ubiquitous, especially with the growing popularity of online social media and e-commerce, are signed networks, which not only have positive links, but also allow for the construction of negative links. For example, negative links can be used to represent distrust, a mechanism to warn others of a potential ``scammer'' in online e-commerce, or in social networks they can represent the connections with our foes (or blocked users). 

Previous work and theories have primarily focused on unipartite signed networks, which are networks that have a single node type and signed links are able to connect any two nodes in the network. However, a common form of signed networks that have primarily been overlooked -- signed bipartite networks. These networks have two sets of nodes and links are only able to be formed between nodes of different types. Actually, signed bipartite networks appear across multiple domains. For example, in e-commerce, a signed bipartite network can be constructed between buyers and sellers in multi-vendor marketplaces when the users are asked to rate the other after each transaction and helpfulness ratings from users to reviews can be naturally denoted as a signed bipartite network. Another important application of signed bipartite networks is from the political science domain, more specifically, we observe that indeed the United States Congress is inherently a signed bipartite network formed from the representatives and the bills they have voted on (where the "Yea" and "Nay" votes can be represented as positive and negative links, respectively)~\cite{derr2018congress,derr2019congress}. In addition, many online systems, such as Netflix and YouTube, adopt ``thumbs-up" or ``thumbs-down" rating systems that can also be formulated as signed bipartite networks. 

Although there have been works focused on unsigned bipartite networks, these methods are lacking the capability to handle the further complexities of negative links. Similarly, methods developed for unipartite signed networks might not be applicable when having the two node types or limiting the possible connections in the network. For example, a fundamental theory that explains the social phenomena of the link structure in signed network analysis is balance theory~\cite{heider1946attitudes,Cartwright-Haray1956}. It suggests that a cycle in signed networks with an even number of negative links is balanced, which is typically stated as ``a friend of my friend is my friend'' while an ``enemy of my friend is my enemy''. In unipartite signed networks balance theory has been extensively applied on signed triangles (i.e., the smallest undirected cycle) across various real-world networks to obtain better performance across modeling~\cite{ludwig2007evolutionary,vukavsinovic2014modeling,derr2018signed}, measuring~\cite{traag2010exponential,bonacich2004calculating,Wu-etal2016,Shahriari-Jalili2014}, and mining applications~\cite{kumar2016edge,chiang2011exploiting,Jung2016srwr,anchuri2012communities} . However, in signed bipartite networks it is fundamentally impossible to have any triangles while having the two different types of nodes. Therefore it is important to understand balance theory in signed bipartite networks and its possibility to enhance applications, due to the prevalence of signed bipartite networks. Thus, dedicated efforts are desired for signed bipartite networks in additional to unipartite signed networks and unsigned bipartite networks.

In this paper, we perform an initial investigation of balance theory in undirected signed bipartite networks. As aforementioned, balance theory has been utilized to advance numerous tasks in unipartite signed networks and sign prediction is the one being benefited most. Hence, we then investigate how to utilize balance theory to boost sign prediction in  signed bipartite networks. This paves the way for using balance theory for other network analysis tasks in signed bipartite networks. The main contributions of the paper are summarized as the following:
\begin{itemize}
\item We conduct the first comprehensive analysis and validation of balance theory for signed bipartite networks; 
\item We leverage balance theory for the construction of multiple sign prediction methods. 
\item We perform experiments on three real-world signed bipartite network datasets to understand balance theory and sign prediction in signed bipartite networks.
\end{itemize}

The remainder of this paper is organized as follows. Our signed bipartite datasets are introduced along with our analysis and investigation of extending balance theory to signed bipartite networks in the form of signed butterflies in Section~\ref{sec:data_analysis}. Then, in Section~\ref{sec:model} we present numerous sign prediction methods for signed bipartite networks based on the signed butterflies. Next we perform experiments for predicting the sign of missing links using our proposed methods in Section~\ref{sec:experiments}. Related work is presented in Section~\ref{sec:related_work} and briefly discussed. Finally, conclusions and future work are given in Section~\ref{sec:conclusion}.

\section{Balance Theory in Undirected Signed Bipartite Networks}\label{sec:data_analysis}

In this section, we will introduce the datasets we have collected for this study. Thereafter we discuss balance theory from a general signed network perspective, then we validate its applicability in signed bipartite networks, and perform a preliminary analysis on our datasets; but first, we introduce the definitions and notations. 

Consider an undirected signed bipartite network, $\mathcal{G} = $ $(\mathcal{U}_B, \mathcal{U}_S,$ 
$\mathcal{E}^+, \mathcal{E}^-)$, where $\mathcal{U}_B = \{b_1, b_2, \dots , b_{n_B}\}$ and $\mathcal{U}_S = \{s_1, s_2, \dots ,s_{n_S}\}$ represent two mutually exclusive sets of homogeneous nodes with $n_B$ and $n_S$ representing the number of nodes for each set, respectively. $\mathcal{E}^+ \subset \mathcal{U}_B \times \mathcal{U}_S$ and  $\mathcal{E}^- \subset \mathcal{U}_B \times \mathcal{U}_S$ represent the sets of positive and negative edges, respectively, between the two sets of nodes $\mathcal{U}_B$ and $\mathcal{U}_S$. We let $\mathcal{E} = \mathcal{E}^+ \cup \mathcal{E}^-$ be the set of all edges where $\mathcal{E}^+ \cap \mathcal{E}^- = \emptyset$, in other words, two nodes cannot have both a positive and negative edge between them. We use $\mathbf{B} \in \mathbb{R}^{n_B \times n_S}$ to represent the undirected signed bipartite biadjacency matrix of $\mathcal{G}$, where $\mathbf{B}_{ij} = 1, -1, $ or 0, when there exists a positive, negative, or no link between $b_i$ and $s_j$. We further summarize the major notations used throughout this work in Table~\ref{tab:notations}. 

\subsection{Signed Bipartite Networks}\label{sec:datasets}
We have collected three signed bipartite networks for this study. The first signed bipartite network is from the e-commerce website Bonanza\footnote{http://www.bonanza.com}. Bonanza is similar to eBay\footnote{http://www.ebay.com} and Amazon Marketplace\footnote{http://www.amazon.com} in that users create an account for which they can buy or sell various goods. After a buyer purchases a product from a seller, both are able to provide a rating about the other along with a short comment. At the time of collection, Bonanza was using a rating scale of ``Positive'', ``Neutral'', and ``Negative'' to rate another user after a transaction. For representing the buyers and sellers, we use $\mathcal{U}_B$ and $\mathcal{U}_S$, respectively.

\begin{table}
\begin{center}
\caption{Notations.}
\vspace{-0.1in}
\label{tab:notations}
\begin{tabular}{|l|l|}
	\hline
	Notations & Descriptions\\
	\hline
	$\mathbf{B}$ & Undirected signed bidjacency matrix\\
    $\mathbf{P}_B$ & Adjacency matrix for the $\mathcal{U}_B$-projection network \\
    $\mathbf{P}_S$ & Adjacency matrix for the $\mathcal{U}_S$-projection network \\
    $\mathbf{A}$ & Adjacency matrix constructed from $\mathbf{B}, \mathbf{P}_B, $ and $\mathbf{P}_S$ \\
    $\mathbf{U}$ & Low-dimensional representation of nodes in $\mathcal{U}_B$\\
    $\mathbf{V}$ & Low-dimensional representation of nodes in $\mathcal{U}_S$\\
	\hline
\end{tabular}
\vspace{-0.05in}
\end{center}
\end{table}

\begin{table}[]
\centering
\caption{Statistics on Signed Bipartite Networks.}\label{tab:datasets_stats} 
\vspace{-0.1in}
\begin{tabular}{|l|c|c|c|}
	 \hline
    & Bonanza & U.S. Senate & U.S. House  \\ \hline
	$n_B = |\mathcal{U}_B|$ & 7,919 & 1,056 & 1,281  \\
    $n_S = |\mathcal{U}_S|$ & 1,973 & 145  & 515  \\ \hline 
    $|\mathcal{E}| = |\mathcal{E}^+| + |\mathcal{E}^-|$ & 36,543 & 27,083 & 114,378 \\
    \% Links Positive & 97.98\% & 55.31\% & 53.96\% \\
    \% Links Negative & 2.02\% & 44.69\% & 46.04\% \\ \hline
	Density of $\mathbf{B}$ & $2.339 \times 10^{-3}$ & 0.1769 & 0.1734 \\  \hline 
\end{tabular}
\vspace{-2.5ex}
\end{table}

\begin{table*}[h!]\small

    \setlength{\extrarowheight}{1.5pt}
    \setlength\tabcolsep{5pt}

\caption{Signed Butterfly Statistics on Signed Bipartite Networks. 
}\label{tab:signed_butterfly_analysis}
\vspace{-2.75ex}
\begin{tabular}{?{0.5mm}l?{0.5mm}c|c|c|c?{0.5mm}c|c|c|c?{0.5mm}c|c|c|c?{0.5mm}} 
		\Xhline{2\arrayrulewidth}
 \hline

    \multirow{2}{2.65cm}{\begin{tabular}{@{}c@{}}Signed Butterfly \\ Isomorphism Classes \end{tabular}} & \multicolumn{4}{c?{0.5mm}}{\textbf{Bonanza}} & \multicolumn{4}{c?{0.5mm}}{\textbf{U.S. Senate}} & \multicolumn{4}{c?{0.5mm}}{\textbf{\begin{tabular}{@{}c@{}}U.S. House of \\ Representatives  \end{tabular}}}\\
	\cline{2-13}	
	& Count & \textbf{\%} & $E\%$ & $s$	& Count & \textbf{\%} & $E\%$ & $s$	& Count & \textbf{\%} & $E\%$ & $s$  \\	
	\Xhline{2\arrayrulewidth}
	\hline
		$(A)~~(+,+,+,+)$	&	2554388	&	0.986	&	0.922	&	386	&	13404168&	0.262	&	0.094	&	4142	&	227660420	&	0.244	&	0.085	&	17459	\\ \hline
        $(B)~~(+,-,-,+)$	&	3830	&	0.001	&	7.8e-04	&	40	&	5595440	&	0.110	&	0.122	&	-277	&	103731010	&	0.111	&	0.123	&	-1137	\\ \hline
		$(C)~~(+,+,-,-)$	&	726	&	2.8e-04	&	7.8e-04	&	-29	&	9404006	&	0.184	&	0.122	&	1349	&	173875858	&	0.186	&	0.123	&	5843	\\ \hline
        $(D)~~(+,-,+,-)$	&	456	&	1.7e-04	&	7.8e-04	&	-35	&	5537080	&	0.108	&	0.122	&	-302	&	101409932	&	0.109	&	0.123	&	-1368	\\ \hline
        $(E)~~(-,-,-,-)$	&	20	&	7.7e-06	&	1.7e-07	&	30	&	6815324	&	0.133	&	0.040	&	3414	&	137478104	&	0.147	&	0.045	&	15104	\\ \hline
\textbf{Balanced} 	&	2559420	&	\textbf{0.988}	&	0.924	&		&	40756018	&	\textbf{0.797}	&	0.500	&		&	744155324	&	\textbf{0.797}	&	0.500	&		\\ 	\Xhline{2\arrayrulewidth}
 \hline
		$(F)~~(+,+,+,-)$	&	30685	&	0.012	&	0.076	&	-390	&	6225745	&	0.122	&	0.302	&	-2811	&	109763190	&	0.118	&	0.289	&	-11565	\\ \hline
        $(G)~~(+,-,-,-)$ &	100	&	3.9e-05	&	3.2e-05	&	2	&	4118075	&	0.081	&	0.197	&	-2099	&	79053742	&	0.085	&	0.210	&	-9430	\\ \hline
\textbf{Unbalanced}	&	30785	&	\textbf{0.012}	&	0.076	&		&	10343820	&	\textbf{0.203}	&	0.500	&		&	188816932	&	\textbf{0.203}	&	0.500	&		\\
		\Xhline{2\arrayrulewidth}
 \hline
\end{tabular}
\end{table*}

The next two datasets are representing the role call votes combined from the 1st to 10th United States Congress. More specifically, we collected two separate datasets\footnote{https://www.govtrack.us/data/}; one for the U.S. Senate and the other for the U.S. House of Representatives (which we will refer to as U.S. House). In each of these datasets we represent the bills that were voted by the set $\mathcal{U}_B$ and the senators or representatives by $\mathcal{U}_S$. If a congressperson voted ``Yea'' or ``Nay'' for the bill, then we represent these as positive or negative links between them, respectively, and leave the connection missing otherwise.

Note that for simplicity throughout the rest of this work we will refer to the nodes in $\mathcal{U}_B$ as ``buyers'' and those in $\mathcal{U}_S$ as ``sellers''. In Table~\ref{tab:datasets_stats} we report some basic statistics of our three collected datasets. We note that in the Bonanza dataset there is a significant imbalance between the number of positive and negative links as compared to the two U.S. Congress datasets. Although these datasets are representing vastly different real-world social structures, we next investigate balance theory~\cite{Cartwright-Haray1956,heider1946attitudes} to the signed bipartite network setting.

\subsection{Signed Butterflies in Bipartite Networks}\label{sec:signed_butterflies_analysis}
\begin{figure}[h]
\begin{center}
\includegraphics[scale=0.25]{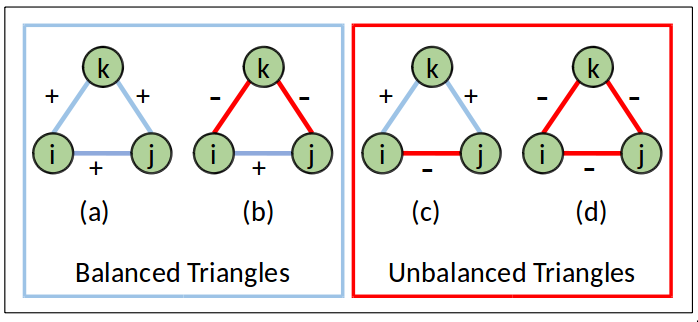} 
\end{center}
\vspace{-2ex}
\caption{Undirected Signed Triangle Isomorphism Classes.\label{fig:balance_triangles}}
\vspace{-1.5ex}
\end{figure}

In signed networks one of the most fundamentally studied social theories is balance theory~\cite{Cartwright-Haray1956,heider1946attitudes}, which discusses the settings in signed networks that are socially ``balanced'' (i.e., stable), and those that are more likely to change (to be balanced) due to the social tensions involved in maintaining ``unbalanced'' and seemingly unnatural connections. In recent signed network analysis works balance theory is usually investigated and then applied towards many tasks\cite{Wang-etal2017,leskovec2010signed,tang2016survey}, but almost always in the form of triangles (or cycles of length 3) in a unipartite signed network. As seen in Figure~\ref{fig:balance_triangles}, there are four possible configurations between the three nodes. We can further observe in Figure~\ref{fig:balance_triangles} that triangles (a) and (b) are balanced (due to having an even number of negative links), while (c) and (d) are unbalanced. Nevertheless, as previously mentioned, since there are no triangles in signed bipartite networks and they have two different node types, it is unknown whether balance theory is still applicable towards a bipartite setting. 

In this subsection, we will therefore introduce how we plan to extend the usage of balance theory to the smallest signed cycles (i.e., butterflies) in undirected signed bipartite networks. Next we investigate and present our initial analysis of these signed butterflies in three real-world signed bipartite networks.

\subsubsection{Signed Butterfly Isomorphism Classes}
In unsigned bipartite networks, one commonly investigated structure is that of a ``butterfly''~\cite{aksoy2017measuring,sanei2018butterfly}, which is a cycle of length 4. 
More formally, a butterfly is the simplest cohesive higher-order structure and also a complete biclique. 
Thus, this provides the most natural structure to investigate as a possible extension for balance theory in signed bipartite networks.

\begin{figure}[h]
\begin{center}
\includegraphics[scale=0.38]{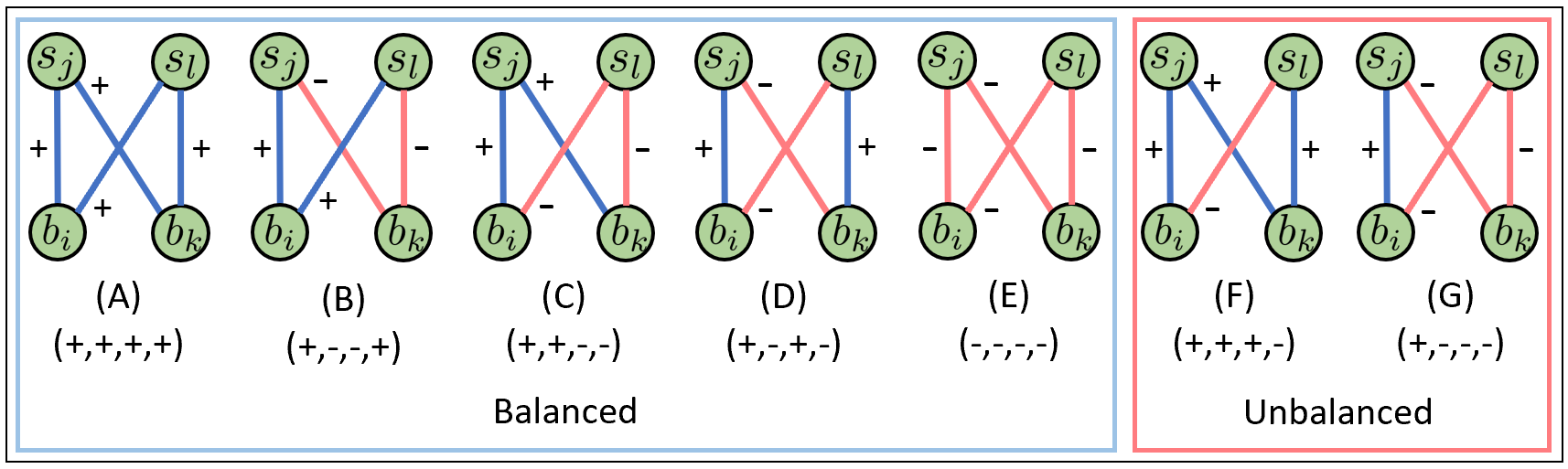} 
\end{center}
\vspace{-2ex}
\caption{Undirected Signed Butterfly Isomorphism Classes.\label{fig:signed_butterflies}}
\vspace{-1.5ex}
\end{figure}

Just as there are different types of signed triangles, there are different types of signed butterflies. In Figure~\ref{fig:signed_butterflies} we present the 7 non-isomorphic undirected signed butterflies. Note that there are five that adhere to balance theory while only two are categorized as unbalanced. We use the notation $(*,*,*,*)$ to denote a signed butterfly isomorphism class that represents the links between the buyers and sellers $(b_i, s_j, b_k, s_l)$ (in that order with the last sign connecting $s_l$ and $b_i$). The simplest of types are $(+,+,+,+)$ and $(-,-,-,-)$, which denote the classes having all positive or all negative links, respectively, and both are balanced due to having an even number of negative links (and can be seen in Figures~\ref{fig:signed_butterflies}(A) and ~\ref{fig:signed_butterflies}(E), respectively). We can interpret the (+,+,+,+) class as the situations where two buyers have bought from the same two sellers and the sentiment amongst them across the four purchases was positive. Next, we have $(+,+,+,-)$ and $(+,-,-,-)$, which are the two unbalanced classes of signed butterflies (since they have an odd number of negative links). In Figure~\ref{fig:signed_butterflies}(F) we have the signed butterfly isomorphism class that encompasses all the signed butterflies with a single negative link. We can observe that no matter where this single negative link is placed, we always have one buyer with two positive links, one buyer with a positive and negative link, and similar structure for the two sellers. The isomorphism class $(+,-,-,-)$ can be seen as the complement (if defined as swapping link signs in a signed network) of the class $(+,+,+,-)$ and defined in a similar way, but with swapping the positive and negative links in the definition. This leaves the signed butterflies having two positive and two negative links, of which we have three isomorphism classes. In Figure~\ref{fig:signed_butterflies}(D) we see the class $(+,-,+,-)$ is used to represent signed butterflies where all buyers and sellers have one positive and one negative link in their cycle. When one of the buyers has two positive links, while the other buyer has two positive links, we observe in Figure~\ref{fig:signed_butterflies}(B) that both sellers have a single positive and single negative link, and define the isomorphism class of $(+,-,-,+)$. Finally, the last type of signed butterfly has both buyers connected positively to one seller, and negatively to the other, which we represent as the class $(+,+,-,-)$ shown in Figure~\ref{fig:signed_butterflies}(C).

\subsubsection{Signed Butterfly Analysis}
In Table~\ref{tab:signed_butterfly_analysis} we report our analysis after counting the number of signed butterflies for each isomorphism class as shown in Figure~\ref{fig:signed_butterflies}. We further calculated the percentage each isomorphism class takes up of the total signed butterfly count in each dataset (given in column ``$\%$''). Next, we analyzed the significance of these signed butterflies being found in signed bipartite networks and wanted to test whether they are overrepresented or underrepresented. Remember, balance theory would suggest that balanced isomorphism classes (A) through (E) should appear frequently while (F) and (G) (being unbalanced) should appear less frequently. To quantify this, extending the approach taken in~\cite{leskovec2010signed}, we calculate ``$E\%$'' as the expected percentage of total signed butterflies to fall into the given isomorphism class when randomly reassigning the positive and negative signs to the signed bipartite network. In other words, for example, ``$E\%$'' for the isomorphism class $(+,-,-,-)$ is calculated by ${4 \choose 1} \Bigl( (|\mathcal{E}^+|/|\mathcal{E}|)\times (|\mathcal{E}^-|/|\mathcal{E}|)^3 \Bigr)$, since there are 4 permutations of having a single positive link in a signed butterfly in class $(+,-,-,-)$ and the probability of each link appearing in a signed network with randomly assigned link signs would be the independent probabilities of having a single positive link (i.e., $|\mathcal{E}^+|/|\mathcal{E}|$) and three negative links (i.e., $|\mathcal{E}^-|/|\mathcal{E}|$). Finally, the value ``$s$'' is used to denote the number of standard deviations the actual count differs from our calculated expected number (based on ``$E\%$'') for each signed butterfly type and just as in~\cite{leskovec2010signed}, a positive (or negative) ``$s$'' value signifies appearing significantly more (or less) than expected.

We first observe that the large majority of signed butterflies in our three signed bipartite networks are indeed balanced. Furthermore, they are significantly more balanced than expected based on the link sign ratio in the given network (i.e., comparing columns ``$\%$'' and $E\%$). The second observation is that all unbalanced signed butterflies across the three datasets are significantly underrepresented, except for the $(+,-,-,-)$ butterflies in Bonanza, where it shows a minimal over representation. 
Similarly, across all datasets the $(+,+,+,+)$ and $(-,-,-,-)$ signed butterflies are significantly overrepresented, 
further strengthening the applicability of balance theory in signed bipartite networks. However, the 
isomorphism classes involving two positive and two negative links appear to not always be found overrepresented. For example, the class where all buyers and sellers have one positive and one negative link, i.e., $(+,-,+,-)$, is less commonly found than expected across all three datasets.

In summary, our findings suggest that: 1) we can use signed butterflies to extend balance theory for signed bipartite networks; and 2) signed bipartite networks adhere to balance theory when defined in terms of signed butterflies, thus making them applicable to advance numerous tasks in signed bipartite networks.

\subsection{Signed Caterpillars in Bipartite Networks}
In this subsection, we discuss the notion of ``signed caterpillars''
, which we denote as paths of length 3 that are missing just one link to becoming a signed butterfly.  

A signed caterpillar can take on one of eight different forms, since it is composed of three links being ether positive or negative. Note that all caterpillar types have the potential to be transformed into a signed butterfly (i.e., closed into a cycle of length 4) that is either balanced or unbalanced. If a signed caterpillar contains an even number of negative links, we refer this as a ``balanced path'' and balance theory would suggest a positive (or negative) link transforming it into a balanced (or unbalanced) signed butterfly. Similarly, we define an a signed caterpillar as an ``unbalanced path'' when having an odd number of negative links and balance theory would suggest a negative (or positive) link to close into a balanced (or unbalanced) signed butterfly.

\section{Sign Prediction for Undirected Signed Bipartite Networks}\label{sec:model}

With the aforementioned definitions and notations, we formally define the problem of sign prediction in undirected signed bipartite networks as the following:

\textit{Given an undirected signed bipartite network $\mathcal{G} = (\mathcal{U}_B, \mathcal{U}_S, \mathcal{E}^+, \mathcal{E}^-)$ represented as a biadjacency matrix $\mathbf{B} \in \mathbb{R}^{|\mathcal{U}_B| \times |\mathcal{U}_S|}$, we seek to predict the signs of no link pairs $(b_i, s_j) \in \{\mathcal{U}_B \times \mathcal{U}_S\} \backslash \{ \mathcal{E}^+ \cup \mathcal{E}^- \}$.} 

Sign prediction in signed networks has been previously studied~\cite{ chiang2011exploiting, javari2014cluster, zhang2013characterization,papaoikonomou2014edge,leskovec2010predicting}. However, in the signed bipartite setting, many of these methods are no longer applicable, since there are no triangles. 
In Section~\ref{sec:signed_butterflies_analysis}, we validated that the large majority of signed butterflies in signed bipartite networks are balanced. Methods for predicting link signs in unipartite signed networks can be categorized into three main groups: 1) supervised methods; 2) low-rank approximation methods; and 3) propagation based methods. Therefore we develop a representative sign prediction method specific to signed bipartite networks from each group. More specifically, we propose: 1) a supervised classification method that uses signed caterpillars/butterflies; 2) extend a low-rank modeling method to ensure the predicted signs favor creating more balanced signed butterflies; and 3) a random walk based approach that integrates one-mode projection networks for $\mathcal{U}_B$ and $\mathcal{U}_S$ constructed using balance theory. 

\subsection{Signed Caterpillars Based Classifier}
One common approach towards predicting links or link signs in both signed and unsigned networks is to frame the task in terms of a supervised classification problem~\cite{leskovec2010predicting,chiang2011exploiting,al2006link,lu2010supervised}. Here we extend the idea to the signed bipartite setting by formulating the problem of predicting the sign between a buyer $b_i$ and a seller $s_j$ by extracting features from either the individuals (i.e., their positive and negative degrees) or local neighborhood features based on balance theory (i.e., signed caterpillars).

To train our model we construct a training dataset consisting of known signed links (between a buyer and seller). Then, after having a trained model, we can extrapolate what we learned from the training data to predict a positive or negative sign for an unknown buyer and seller pair. More specifically, we use a logistic regression model following the prediction on directed signed unipartite networks work in~\cite{leskovec2010predicting}. 

{\bf Feature Extraction.} The two different sets of features we evaluate are either based on the two nodes degree distributions or information about how many signed caterpillars they are the two endpoints of (i.e., they would be the buyer and seller connection transforming the signed caterpillar to a balanced or unbalanced signed butterfly). Thus, the feature vector ${\bf x}^{d}_{ij}$ for the pair $(b_i,s_j)$ includes the the positive and negative degrees for both $b_i$ and $s_j$. 
In comparison, ${\bf x}^{sc}_{ij}$ contains the counts for each of the 8 possible signed caterpillars that have $b_i$ and $s_i$ as the endpoints.  
The expectation is that the features ${\bf x}^{sc}_{ij}$ will be more informative than those of ${\bf x}^{d}_{ij}$ because they would provide a vast amount of informaiton as to whether their link sign is likely to be positive or negative according to balance theory when considering the types of signed butterflies that would be constructed. This is in comparison to only using the degrees in a method similar in nature to a signed preferential attachment model with ${\bf x}^{d}$. We denote the supervised classifiers that use ${\bf x}^d_{ij}$ and ${\bf x}^{sc}_{ij}$ as SCd and SCsc, respectively.

\subsection{Low-Rank Sign Prediction}
In recent years the low-rank matrix factorization approaches have been gaining popularity for numerous applications involving link related network predictions~\cite{menon2011link,menon2011link, tang2016recommendations, hsieh2012low}. Although some of these works have focused on signed networks~\cite{hsieh2012low,forsati2015pushtrust,tang2016recommendations}, none are structured to select link signs that would explicitly push towards more signed butterflies being balanced in signed bipartite networks. Thus, we first introduce a basic matrix factorization approach to model the signed bipartite network using the biadjacency matrix ${\bf B}$. Then we introduce how we can successfully modify this model through the inclusion of additional pairs of buyers and sellers derived from suggested implicit signed links that would construct the most balanced signed butterflies with the suggested link sign. 

\subsubsection{Basic Matrix Factorization Model}
The set of existing edges in ${\bf B}$ are denoted in the set $\mathcal{E} = \{(b_i,s_j) | {\bf B} \neq 0\}$. In terms of the link sign prediction task we would like to discover two latent matrices ${\bf U} = [{\bf u}_1, {\bf u}_2, \dots, {\bf u}_{n_B}] \in \mathbb{R}^{d \times n_B}$ and ${\bf V} = [{\bf v}_1, {\bf v}_2, \dots, {\bf v}_{n_S}] \in \mathbb{R}^{d \times n_S}$ of dimension $d$ for the set of buyers and sellers, respectively, to solve the following optimization problem:
\begin{align}\label{eq:basic-mf}
\min_{{\bf U},{\bf V}} & \sum\limits_{(b_i,s_j) \in \mathcal{E}} \max \Big( 0, 1 -  {\bf B}_{ij} ({\bf u}_i^\top {\bf v}_j) \Big)^2 + \lambda \Big( |{\bf U}|^2_F + |{\bf V}|^2_F \Big) 
\end{align}

\noindent where ${\bf u}_i^\top {\bf v}_j$ is used to model the link sign between buyer $b_i$ to seller $s_j$. Note that when the real link sign (i.e., ${\bf B}_{ij}$) and the predicted link sign (i.e.,$ {\bf u}_i^\top {\bf v}_j$) are of the same sign (i.e., both positive or both negative) then ${\bf B}_{ij} ({\bf u}_i^\top {\bf v}_j)$ is positive, and if over 1 then there is no loss. However, when the real and predicted values have differing signs then there is a higher loss value associated to drive the minimization during the training process. Following the work in~\cite{hsieh2012low} we use Stochastic Gradient Descent (SGD) to minimize the objective in Eq.~(\ref{eq:basic-mf}). 

This allows us to then utilize the learned low-dimensional representations for each buyer and seller to predict the sign of unknown buyer and seller pairs. However, although this model is effectively learning a representation that can accurately predict the existing links, it does not explicitly control whether the signs of non-existing links are actually going to predict link signs that adhere to balance theory (i.e., having more signed butterflies balanced than unbalanced). Therefore we denote this method simply as MF. Next, we will present an extension to this basic framework to further ensure more signed butterflies between the missing links are balanced. 

\subsubsection{Matrix Factorization with Balance Theory}
As previously discussed, the aforementioned basic matrix factorization approach given in Eq.~(\ref{eq:basic-mf}) does not explicitly enforce the non-existing link signs to favor balanced relationships. Instead it can only focus on learning low-dimensional representations for each buyer and seller such that the model minimizes the error on predicting the existing link signs. The approach we have selected is to further encourage the model learning link signs for buyer and seller pairs that currently do not exist in the signed bipartite network, but would convert many signed caterpillars into balanced signed butterflies if they were to exist. 

The first step is calculating whether balance theory would suggest a positive or negative link for each buyer and seller pair ($b_i$,$s_j$), that currently do not have a link between them, based on the types of signed caterpillars they're jointly involved in and the endpoints of.

\begin{theorem}\label{theorem:low-rank-additional-links}
Given a signed undirected biadjacency matrix ${\bf B}$, then the matrix ${\bf \hat{S}} = {\bf B}{\bf B}^T{\bf B} \odot {\bf \overline{B}}$ is such that $sign({\bf \hat{S}}_{ij})$ suggests the sign of a non-existent link in ${\bf B}$ that would result in a net gain of $|{\bf \hat{S}}|$ additional balanced signed butterflies created (after subtracting the number of potential unbalanced signed butterflies created simultaneously) if the suggested signed link were to be added between $b_i$ and $s_j$, where we define ${\bf \overline{B}}$ as ${\bf \overline{B}}_{ij} = 0$ if \hspace{0.15ex} ${\bf B}_{ij} \neq 0$ and ${\bf \overline{B}}_{ij} = 1$ when ${\bf B}_{ij} = 0$. 
\end{theorem}
 \vspace{-2ex}
 \begin{proof}
 If we let $\mathbf{A} = \Bigl[ \begin{smallmatrix}{\bf 0} & {\bf B} \\ {\bf B}^T & {\bf 0}\end{smallmatrix}\Bigr]$ be the adjacency matrix in $\mathbb{R}^{|\mathcal{U}| \times |\mathcal{U}|}$. We can observe that $\mathbf{A}^3 = \Bigl[ \begin{smallmatrix}{\bf 0} & {\bf B}{\bf B}^T{\bf B} \\ {\bf B}^T{\bf B}{\bf B}^T & {\bf 0}\end{smallmatrix}\Bigr]$. We note that in~\cite{derr2018relevance} it has been shown ${\bf A}^l = {\bf M}^l_B - {\bf M}^l_U$, where ${\bf M}^l_B, {\bf M}^l_U  \in \mathbb{R}^{|\mathcal{U}| \times |\mathcal{U}|}$ store the number of balanced and unbalanced paths of length $l$, respectively, between all pairs of nodes in a signed network represented as ${\bf A}$. Thus, since ${\bf A}^3_{ij} = \bigl[{\bf B}{\bf B}^T{\bf B}\bigr]_{ij}$ for some buyer $b_i$ and seller $s_j$, we observe that this represents the number of of balanced paths of length 3 subtracted by the number of unbalanced paths of length 3. By definition of a signed caterpillar, if one is a balanced path, then it would suggest a positive link to close to be a balanced signed butterfly, but if it was formed by an unbalanced path it would require the closing link to be negative to form a balanced butterfly. Therefore, it follows that $sign(\bigl[ {\bf B}{\bf B}^T{\bf B} \bigr]) = sign\big({\bf M}^l_B - {\bf M}^l_U \big)$ indeed represents the sign that would promote the creation of more balanced signed butterflies, and similarly for the net gain of balanced butterflies being formed equaling the absolute value of their difference $(i.e, |{\bf M}^l_B - {\bf M}^l_U|)$. It is then easy to extend to only the buyer and seller pairs $b_i$ and $s_j$ in $\bigl[ {\bf B}{\bf B}^T{\bf B} \bigr] \odot \overline{\bf B}$ after taking the element-wise product with ${\bf \overline{B}}$ that zeros out the pairs that have an existing link.
 \end{proof}

Note that ${\bf \hat{S}}$ can also be calculated (sometimes more efficiently) using the following:
{
\[
  {\bf \hat{S}}_{ij} =
  \begin{cases}
  \bigl[{\bf B}{\bf B}^\top{\bf B}\bigr]_{ij} & \text{if ${\bf B}_{ij} = 0$} \\
  0 & \text{otherwise}
  \end{cases}
\]
}

\noindent to avoid using the potentially very dense matrix ${\bf \overline{B}}$ for sparse signed bipartite networks.

Using Theorem~\ref{theorem:low-rank-additional-links} we can construct additional sets $\mathcal{E}^+_i$ and $\mathcal{E}^-_i$ of implicit positive and negative links, respectively, suggested by balance theory that would create the highest net gain of balanced signed butterflies in the signed bipartite network.  
We define these sets as follows: 
\begin{align}
&\hat{\mathcal{E}}^+_i = \{(b_i,s_j) \hspace{0.2ex} | \hspace{0.2ex} {\bf \hat{S}}_{ij} > 0 \text{ and } {\bf \hat{S}}_{ij} \in top_k ( {\bf \hat{S} } ) \} \nonumber \\
&\hat{\mathcal{E}}^-_i = \{(b_i,s_j) \hspace{0.2ex} | \hspace{0.2ex} {\bf \hat{S}}_{ij} < 0 \text{ and } {\bf \hat{S}}_{ij} \in bottom_k ( {\bf \hat{S}} ) \}
\end{align}

\noindent where $top_k ( {\bf \hat{S} } )$ and $bottom_k ( {\bf \hat{S} } )$ are used to denote the $k$ largest and smallest values, respectively, in ${\bf \hat{S}}$.

We formulate our object that incorporates balance theory as follows:
\begin{align}
& \min_{{\bf U},{\bf V}}  \sum\limits_{(b_i,s_j) \in \mathcal{E}} \max \Big( 0, 1 -  {\bf B}_{ij} ({\bf u}_i^\top {\bf v}_j) \Big)^2 + \lambda \Big( |{\bf U}|^2_F + |{\bf V}|^2_F \Big) \nonumber \\
&+  \alpha \sum\limits_{(b_i,s_j) \in \hat{\mathcal{E}}^+_i} \max \Big( 0, 1 -  {\bf \hat{S}}_{ij} ({\bf u}_i^\top {\bf v}_j) \Big)^2 \nonumber \\
&+ \beta \sum\limits_{(b_i,s_j) \in \hat{\mathcal{E}}^-_i} \max \Big( 0, 1 -  {\bf \hat{S}}_{ij} ({\bf u}_i^\top {\bf v}_j) \Big)^2  
\end{align}

\noindent where $\alpha$ and $\beta$ are used to control the level at which we incorporate the modeling of signed butterflies through the inclusion of the implicit positive and negative links, respectively. We again note that these implicit positive and negative links are implied by balance theory by using ${\bf \hat{S}}$, which effectively counts for each node pair $(b_i,s_j)$ what the net gain of total balanced signed butterflies would be once including the link with the suggested sign (according to the majority count of signed caterpillars being of balanced or unbalanced paths of length 3). We denote this matrix factorization method using balance theory as MFwBT.

\subsection{Random Walk Based Sign Prediction}

Typical propagation based methods, such as the random walk with restart~\cite{tong2006fast} have seen many variants and been applied to solve link prediction and ranking related tasks in unsigned unipartite networks. However, signed bipartite networks pose multiple challenges that prevent them from directly using the typical methods. One such problem is that bipartite networks do not have a stationary distribution and thus do not converge~\cite{lovasz1993random}. One 
way of handling this problem in unsigned bipartite networks is considered a ``lazy'' random walk, where the walker will probabilistically stay at the same node. We will later use this method as a comparison against our proposed random walk based method.  
Furthermore as seen in previous sign prediction methods for unipartite signed networks, balance theory is the key component towards obtaining higher performance when predicting the sign of unknown links. Thus, due to our analysis of the signed butterflies, indeed signed bipartite networks are showing high levels of balance and therefore we should also be using balance theory to guide the random walk based method for signed bipartite networks towards a solution having more balanced relations.

Here we present a random walk based approach that integrates the $\mathcal{U}_B$ and $\mathcal{U}_S$ one-mode projection adjacency matrices, which are constructed using balance theory, to aid in handling the issues faced with the bipartite setting, and develop a signed random walk based approach to not only allow a proper transition matrix, but to furthermore have the random walker be promoting balance theory. The first step will be the construction of a signed adjacency matrix $\bf{A}$ based on balance theory, followed by defining a signed transition matrix that can further promote and propagate balanced relations throughout the network.

\begin{figure}[t]
\begin{center}
\includegraphics[scale=0.42]{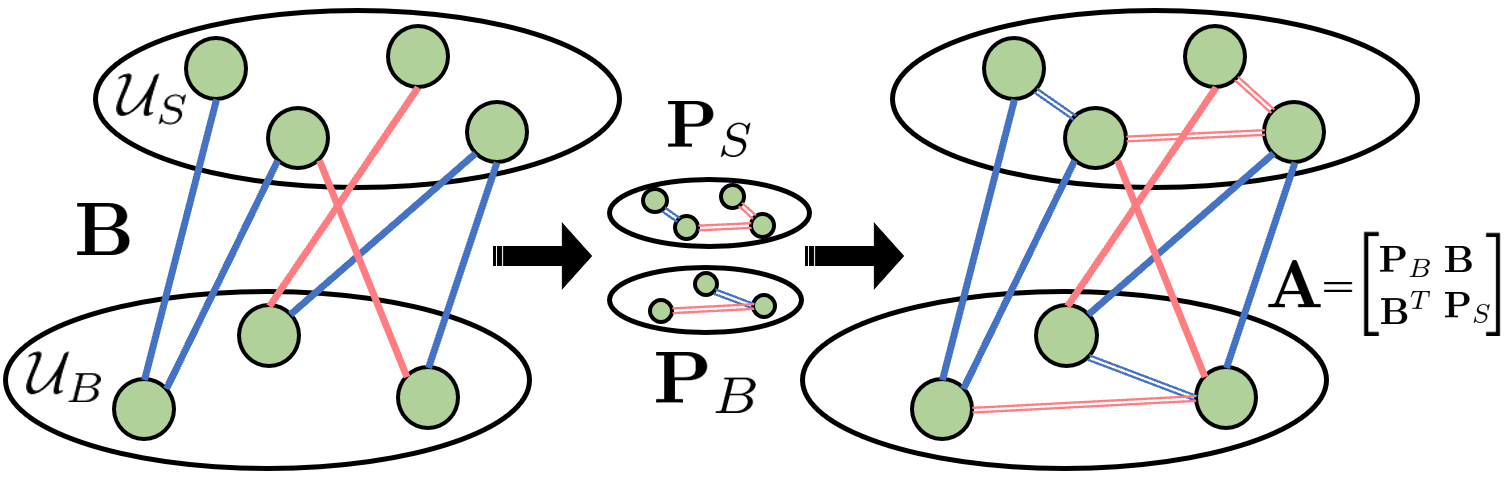}
\end{center}
\caption{High-level intuition of how we construct ${\bf A}$.
\label{fig:sbrw}}
\end{figure}

\subsubsection{Constructing the one-mode adjacency matrices.}
In unsigned bipartite network analysis one-mode projections are typically used for both analysis and aiding to solve various tasks~\cite{gao2017projection,zhou2007bipartite,zweig2011systematic}. They are constructed by creating a projection network that creates implicit connections between nodes of the same type. In terms of our definitions, two one-mode projection networks can be performed, one that connects the buyers in $\mathcal{U}_B$ together amongst themselves and the other for the sellers in $\mathcal{U}_S$ by constructing seller to seller links; these relations can be represented in the adjacency matrices $\mathbf{P}_B \in \mathbb{R}^{|\mathcal{U}_B| \times |\mathcal{U}_B|}$ and $\mathbf{P}_S \in \mathbb{R}^{|\mathcal{U}_S| \times |\mathcal{U}_S|}$. A visual example can be seen in Figure~\ref{fig:sbrw} when going from $\mathbf{B}$ to $\mathbf{P}_S$ and $\mathbf{P}_B$ from left to right through the first arrow.

We note that there is not just one way to discover these implicit connections between pairs of users in the same set, and in fact there are many possible methods for one-mode projections~\cite{zhou2007bipartite,zweig2011systematic}. It has also been studied that using different methods to construct the projection networks can cause drastic changes to the usability and performance~\cite{yildirim2014using}. In wanting to carefully construct these projection networks, we choose to utilize balance theory in the form of signed triangles. Next we will discuss the formation of the adjacency matrix $\mathbf{P}_B$, and a similar process can be followed for constructing $\mathbf{P}_S$ (although we only discuss $\mathbf{P}_B$ here). 

Based on the ideas of common neighbor similarity in unsigned networks, we will possibly connect two buyers $b_i$ and $b_j$ if they have at least one seller in common they are linked to. Let the number of common sellers that $b_i$ and $b_j$ agree upon (in terms of link sign) be denoted as $ns^{A}_{ij}$. Similarly let $ns^{D}_{ij}$ denote the number of sellers these two buyers disagree on in terms of link sign. Then we define $\mathbf{P}_{Bij} = \mathbf{P}_{Bji} = ns^{A}_{ij} - ns^{D}_{ij}$, which we can see is taking the number of sellers they agree upon in terms of signed connections (i.e., both either negatively or positively connected to that seller) and subtracting the number of sellers they disagree on (i.e., the sellers where one buyer has a positive link while the second buyer has a negative connection with that seller). We can now further see the connection between $\mathbf{P}_{Bij}$ and the common neighbor similarity method. 
It is easy to verify that out of all the triangles formed between $b_i$, $b_j$, and the sellers $s_k$ they are commonly linked to, that using the links $\mathbf{B}_{ik}$, $\mathbf{B}_{jk}$ and $\mathbf{P}_{Bij}$, we see that the majority will adhere to balance theory. This is by design since if $na^{A}_{ij} > na^{D}_{ij}$ then $sign(\mathbf{P}_{Bij})$ is positive and closing the $ns^{A}_{ij}$ triangles to be balanced, while the lesser number of $ns^{D}_{ij}$ will close to be unbalanced. Note that a similar argument can be given when $ns^{A}_{ij} < ns^{D}_{ij}$ and if $ns^{A}_{ij} = ns^{D}_{ij}$ then $\mathbf{P}_{Bij} = 0$ and no signed triangles are formed. Ultimately, we construct a parameterized version as follows:
\begin{align}
\mathbf{P}_{Bij} = 
	\left\{
        \begin{array}{ll}
            0 & \quad  \delta_n < ns^{A}_{ij} - ns^{D}_{ij} < \delta_p \\
            ns^{A}_{ij} - ns^{D}_{ij} & \quad \text{otherwise}
        \end{array}
    \right. 
\end{align}

\noindent where $\delta_p$ and $\delta_n$ are used to define thresholds for the necessary magnitude of $ns^{A}_{ij} - ns^{D}_{ij}$ to have a non-zero value in $\mathbf{P}_{Bij}$. This allows us to ignore adding smaller values (e.g., 2), since in some settings having such a small value might not be very significant and thus we might not want to construct a link between $b_i$ and $b_j$. Note that for simplicity we allow $\delta_p$ and $\delta_n$ to be shared for constructing both ${\bf P}_B$ and ${\bf P}_S$.

\subsubsection{Performing the random walk.}
Now having the two projection adjacency matrices ${\bf P}_B$ and ${\bf P}_S$, we can use them to construct an adjacency matrix ${\bf A} \in \mathbb{R}^{|\mathcal{U}| \times |\mathcal{U}|}$, which will be the unibipartite signed network we perform our random walk on, where $\mathcal{U} = \{b_1, \dots b_{n_B}, s_1, \dots s_{n_S}\}$. In Figure~\ref{fig:sbrw} we show the high-level intuition of how to construct ${\bf A}$. First we denote ${\bf \hat{B}}$ as the row normalized biadjacency matrix where ${\bf \hat{B}}_{ij} = {\bf B}_{ij}/\sum\limits_{k}^{}{|{\bf B}_{ik}|}$. We similarly construct row normalized adjacency matrices ${\bf \hat{P}}_B$ and ${\bf \hat{P}}_S$. Now we can formulate ${\bf A}$ as follows:
\begin{align}
{\bf A} = 
\begin{bmatrix}
    {\bf \hat{P}}_B  &  \omega {\bf \hat{B}}      \\
    \omega {\bf \hat{B}}^T  &  {\bf \hat{P}}_S      
\end{bmatrix}
\end{align}

\noindent where $\omega$ is a parameter that can be used to bias the random walker to favor the real links in our signed bipartite network as compared to the implicit links we obtained through the $\mathcal{U}_B$ and $\mathcal{U}_S$ one-mode projection networks. Next we construct a similar row normalized adjacency matrix ${\bf \hat{A}}$ where ${\bf \hat{A}}_{ij} = {\bf A}_{ij}/\sum\limits_{k}^{}{|{\bf A}_{ik}|}$. 

Finally, we utilize ${\bf \hat{A}}$ in a random walk propagation model where we define ${\bf Y}$ to be the matrix holding the inferred link signs as follows:
\begin{align}\label{eq:signed_random_walk}
{\bf Y}_{ij} = \sum\limits_{k}{\bf \hat{A}}_{ik}{\bf Y}_{kj} 
\end{align}

\noindent Next we describe how the above adheres to balance theory in terms of triangles of the adjacency matrix ${\bf A}$. This is because for some $k$ if ${\bf \hat{A}}_{ik}{\bf Y}_{kj} > 0$ it increases ${\bf Y}_{ij}$ ensuring it to be positive, which would be a triangle consisting of either three positives, or two negatives and a positive. Similarly when ${\bf \hat{A}}_{ik}{\bf Y}_{kj} < 0$ we are decreasing ${\bf Y}_{ij}$ and encouraging it to be negative, thus also following balance theory. The closed form solution that includes the restart capability, with probability $(1-c)$, is given to be the following: 
\begin{align}
{\bf Y} = (1-c)({\bf I} - c{\bf \hat{A}})^{-1}
\end{align}

\noindent Note that each signed butterfly involving $b_i, s_j, b_k,$ and $s_l$ in the original network ${\bf B}$ now consists of up to ${4 \choose 3}$ triangles in ${\bf \hat{A}}$. Thus, when we are encouraging balanced triangles here in ${\bf Y}$ this correlates to having balanced signed butterflies in the upper right corner of ${\bf Y}$, which is where we obtain the link sign predictions (i.e., when predicting the sign between $b_i$ and $s_j$ we have use $\hat{{\bf B}}_{ij'}$ where $j'=(n_{B} + j)$. We denote this method as Signed Bipartite Random Walk (SBRW).

For comparison, if we set the two one-mode projection matrices to the identity matrix (i.e., ${\bf P}_B = {\bf P}_S = {\bf I}$) and set $\omega=1$ then Eq.~\ref{eq:signed_random_walk} becomes the equation for a lazy random walk method, which we denote as LazyRW.

\section{Experiment}\label{sec:experiments}
In this section, we empirically evaluate our proposed sign prediction methods for signed bipartite networks that harness balance theory. We seek to answer the following: (1) Does the extended balance theory to signed butterflies in the bipartite setting provide an increase in performance for sign prediction? and (2) How do the proposed methods work/compare? To address these questions we perform experiments to measure the performance for each of the proposed sign prediction methods across three real-world signed bipartite networks. To better understand our methods and the contribution of balance theory, we also follow-up with a parameter sensitivity analysis for the major parameters of our methods. 

\subsection{Experimental Settings}
Here we discuss the settings used for our experiments on sign prediction in signed bipartite networks. As previously discussed in Section~\ref{sec:datasets} we have collected three signed bipartite networks for this study, namely, Bonanza, U.S. Senate, and U.S. House. 
For our sign prediction experiments we have randomly selected 10\% of the links as test, utilized a random 5\% for validation purposes of tuning the hyperparameters of our models, and the remaining 85\% as training for each of our datasets. More specifically, each method is only given access to the signed bipartite network induced from the training links, then, for each edge in the testing set, we compare the ground truth link sign with the link sign the specific method suggests for that undirected pair. For evaluation we use both F1 and Area Under the receiver operating characteristic Curve (AUC), since the positive and negative links are unbalanced especially in the Bonanza dataset. To the best of our knowledge this is the first study of predicting link signs in signed bipartite networks; hence other existing methods either for unipartite signed or unsigned bipartite networks are likely not applicable. The main investigation is two-fold. First, we want to test the applicability of balance theory (based on signed butterflies) to aid in sign prediction. Second, we want to provide insights to guide practical usage of sign predictors with different types of signed bipartite networks. Thus, we only provide a comparison against the methods we have presented in this work.  Our code and data are available at \url{https://github.com/DSE-MSU/signed-bipartite-networks}.

\subsection{Comparison Results}
The results across our three signed bipartite networks in terms of AUC and F1 can be found in Table~\ref{tab:link_prediction_results_auc_f1} and the first observation we make is that there is not one proposed method that outperforms the others across all the datasets.

\begin{table}

    \setlength{\extrarowheight}{3pt}

	\caption{Link Sign Prediction Results in terms of (AUC,F1).}\label{tab:link_prediction_results_auc_f1}
	\centering
	\begin{tabular}{?{0.5mm}c?{0.5mm}c|c|c?{0.5mm}} \Xhline{2\arrayrulewidth} \hline
		\begin{tabular}{@{}c@{}}Sign Prediction \\Method\end{tabular}& Bonanza &U.S. Senate &U.S. House \\ \Xhline{2\arrayrulewidth}  \hline
		SCd	&	(0.553~,~0.959) 	&	(0.638~,~0.654)	&	(0.625~,~0.635)		\\ \hline
        SCsc	&	(0.664~,~0.674) 	&	(0.812~,~0.823)	&	(0.827~,~0.837)		\\ \Xhline{2\arrayrulewidth} \hline
		MF		&	(0.593~,~0.903) 	&	(0.792~,~0.812)	&	(0.831~,~0.846)		\\ \hline
		MFwBT	&   (0.608~,~0.905)    &   (0.814~,~0.827)	&   (0.834~,~0.848)  	\\ \Xhline{2\arrayrulewidth} \hline
		LazyRW		& (0.547 ~,~0.979) & (0.808 ~,~ 0.821)	&	(0.815 ~,~ 0.827) \\ \hline
		SBRW	&	(0.582 ~,~ 0.949)	&	(0.836~,~0.849)	&	 (0.846~,~0.858)	 \\ \Xhline{2\arrayrulewidth} 
\hline
	\end{tabular}
    \vskip -1ex
\end{table}

The second observation we make is that the three methods SCsc, MFwBT, and SBRW, which receive aid in prediction from balance theory when defined using signed butterflies, always perform better than their respective baseline method (i.e., SCd, MF, LazyRW) that only use generic signed network information in terms of AUC and only in two cases the F1 is worse. In the Bonanza dataset we have the SCd and LazyRW outperforming SCsc and SBRW, respectively, in terms of F1 (although performing worse in AUC). The reason for this is the heavy imbalance between the positive and negative links in this dataset, more specifically, almost 98\% of the links are positive, which is generally a setting where the AUC measurement is preferred to understand the performance better. Therefore we can see that to better detect the few negative links comes at the sacrifice of misclassifying some of the positive links, which is why the F1 of SCsc and SBRW is less than SCd and LazyRW, but comes with a significant increase in AUC. In general we observe that in fact the usage of signed butterflies for sign prediction in signed bipartite networks provides a very significant improvement in almost all cases. This fact suggests that we can give a positive answer to our first question -- the usage of balance theory in the form of signed butterflies for sign prediction in signed bipartite networks indeed provides an empirically verifiable improvement.

In the U.S. Senate and U.S. House datasets, for the methods constructed based on intuitions of how to correctly ensure more balanced signed butterflies are being created when predicting missing link signs (i.e, SCsc, MFwBT, and SBRW), we see the low-rank model outperforms the the supervised classifier approach, while the random walk method performs the best (for both AUC and F1).  

However, unlike the two U.S. Congress datasets, in the Bonanza dataset we actually observe the complete opposite behavior (in terms of AUC) for the ranking of methods that utilize the signed butterfly based balance theory. We hypothesize this is due to the heavy class imbalance between the positive and negative links. With this imbalance the SBRW method might be unable to directly handle this setting as the parameters only focus on separating real/implicit and balance/unbalance through $\omega$ and $\delta_p/\delta_n$. Futhermore, if most negative links are involved in balance relationships then actually this would cause even more positive links to be constructed in the two one-mode projection matrices (since two negatives would result in a positive link being created). In comparison, MFwBT is able to more accurately control the ratio of positive to negative implicit links being used in the training procedure (through selecting the size of both $\hat{\mathcal{E}}^+_i$ and $\hat{\mathcal{E}}^-_i$) when extracting them from investigating which links would cause the most signed caterpillars to turn into balance signed butterflies. Also, we note that in our study we fixed $\alpha = \beta$, but this mechanism would further allow MFwBT to balance the contribution of implicit positive and negative links towards learning the most effective representations. Finally, although we see a drastic improvement in terms of AUC for the SCsc method, we also observe this comes at great cost to the F1 measure, and thus this method is just discovering a trade-off of predicting more negative links. This is because we have tuned our logistic regression model to use weights on each training example inversely proportional to the frequency of that link type.

\begin{figure}
\begin{center}
\subfigure[MFwBT (AUC) ]{\label{fig:MFwBT_auc}\includegraphics[scale=0.21]{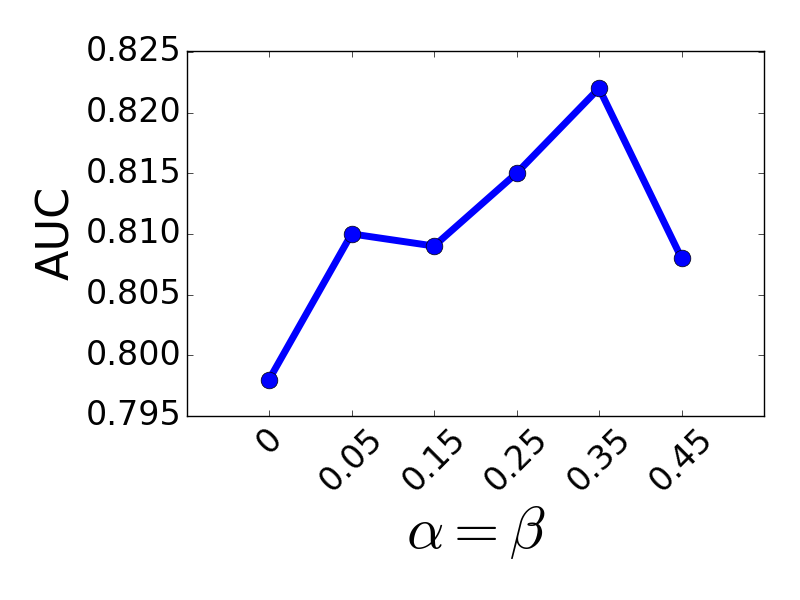}} 
\subfigure[MFwBT (F1)]{\label{fig:MFwBT_f1}\includegraphics[scale=0.2]{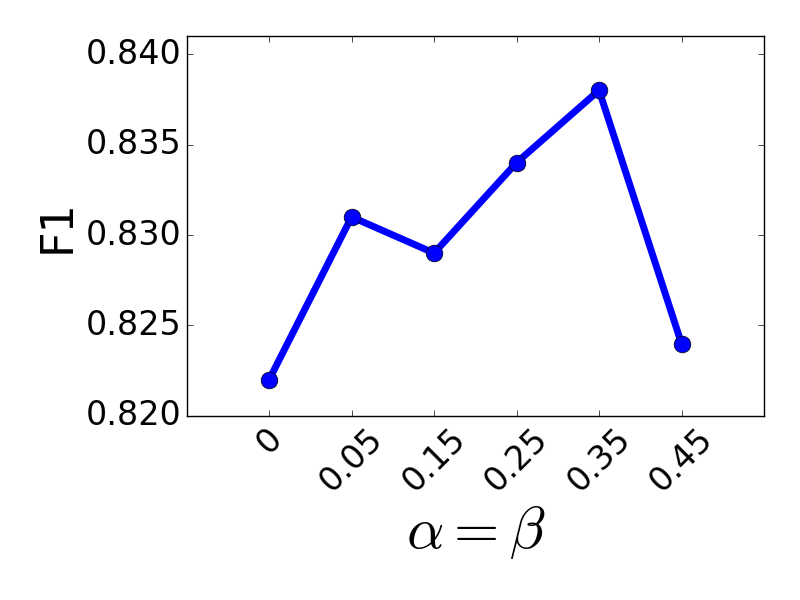}}
\end{center}
\vspace{-3ex}
\caption{Parameter Sensitivity on $\alpha$ and $\beta$ in MFwBT on the U.S. Senate dataset.}\label{fig:MFwBT_alpha_beta}
\vskip -3ex
\end{figure}

\subsection{Parameter Analysis}

Among our three proposed sign prediction methods, the low-rank modeling with balance theory (MFwBT) and random walk (SBRW) methods contain interesting hyperparameters from the perspective of wanting to further understand balance theory in signed bipartite networks.

In our MFwBT method, we discussed that we can control the number of suggested implicit positive and negative links from signed butteflies being included in $\mathcal{E}^+_i$ and $\mathcal{E}^-_i$, respectively. We performed a grid search for both the size of $\mathcal{E}^+_i$ and $\mathcal{E}^-_i$, in the set \{0,1000,10000\}. We discovered that the best setting when considering across the three datasets was having $|\mathcal{E}^+_i|=1000$ and $|\mathcal{E}^-_i|=10000$, which we suspect is due to the class imbalance and having more explicit positive links than negative links. Furthermore, the values of $\alpha$ and $\beta$ were used to control the contribution of training on both positive and negative links suggested based on signed butterflies (i.e., links in $\mathcal{E}^+_i$ and $\mathcal{E}^-_i$), respectively. For simplicity of our analysis we set $\alpha = \beta$ and report the performance on our validation set for the U.S. Senate dataset in Figure~\ref{fig:MFwBT_alpha_beta}. We observe that updating the node representations using suggested signed links (that were selected since they would close the most signed caterpillars into balanced signed butterflies) provides an improvement over not taking balance theory into account (which is when $\alpha=\beta=0$), but care should be taken to not put too much focus on these implicit links. We observe similar findings in our other datasets.

\begin{figure}
\begin{center}
\subfigure[SBRW (AUC)]{\label{fig:SBRW_auc}\includegraphics[scale=0.082]{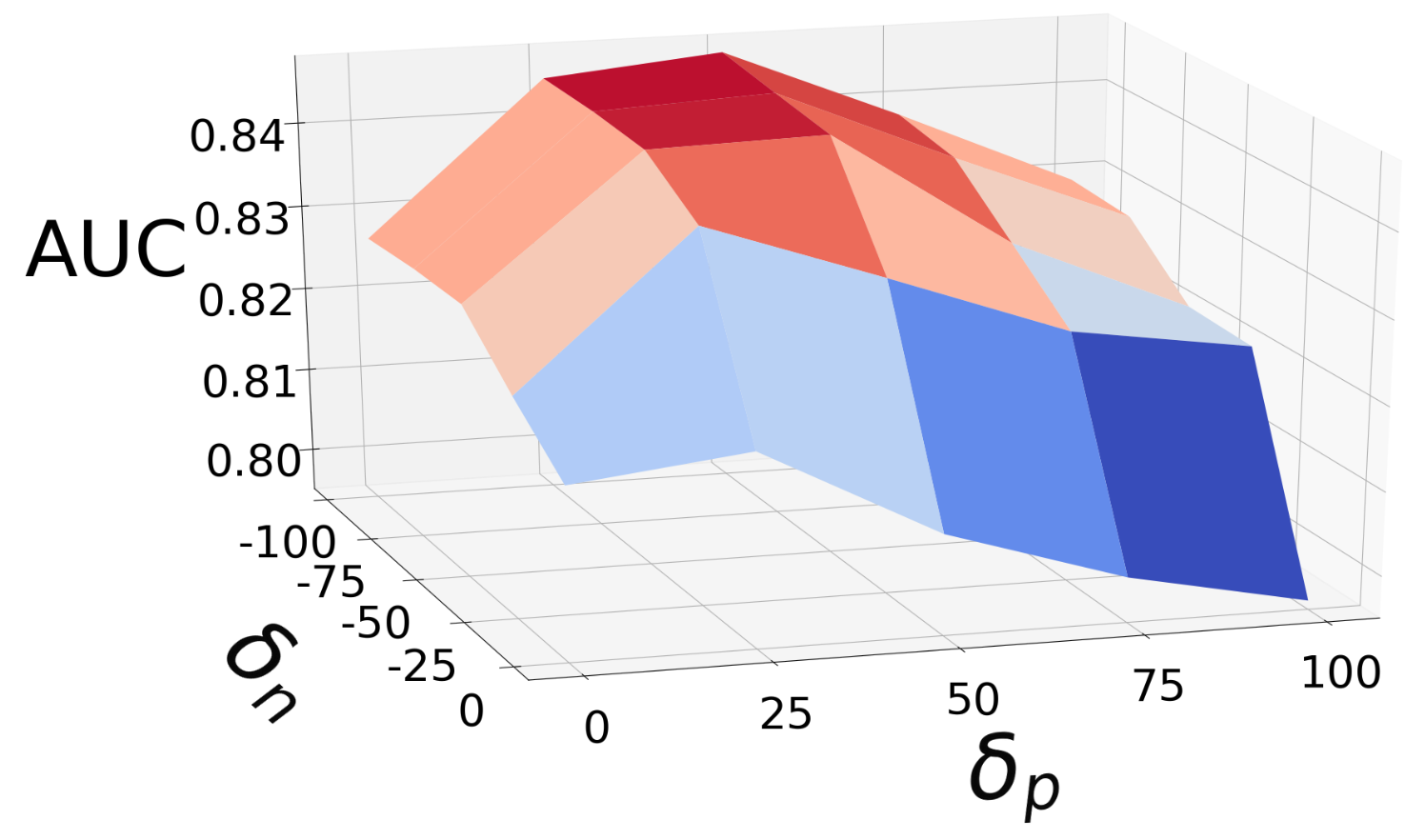}}
\subfigure[SBRW (F1)]{\label{fig:SBRW_f1}\includegraphics[scale=0.082]{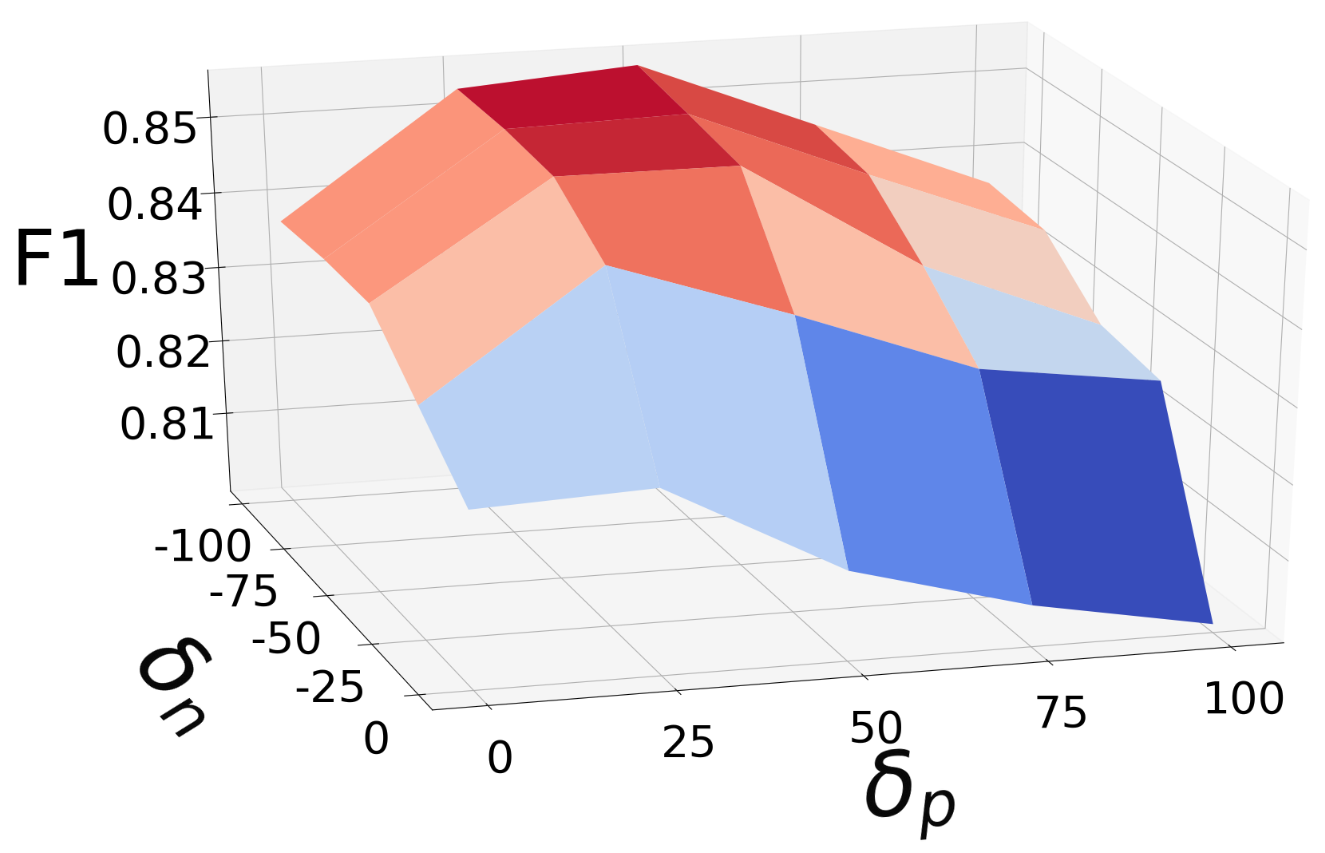}}
\end{center}
\vspace{-3ex}
\caption{Parameter Sensitivity on $\delta_p$ and $\delta_n$ in SBRW on the U.S. House dataset.}\label{fig:SBRW_deltas}
\vskip -2ex
\end{figure}

For our SBRW method, there are two main sets of parameters $\omega$ and the threshold pair $\delta_p$ with $\delta_n$. We varied $\omega$ at a large granularity in the set $\{1,2,3,5\}$, and observed there was not as much significant difference as found in varying $\delta_p$ and $\delta_n$, thus we selected the best on average across the three datasets of $\omega=2$ and fixed this value to investigate the impact of $\delta_p$ and $\delta_n$ on the performance for predicting the missing link signs. 

In Figure~\ref{fig:SBRW_deltas}, for the U.S. House dataset, we varied both $\delta_p$ in the set \{0,25,50,75,100\} and similarly for $\delta_n$ in the set \{0,-25,-50, -75, -100\}. Note that although we saw similar trends across the three datasets, the specific magnitude of $\delta_n$ and $\delta_p$ we needed to tune separately for each dataset due to the average magnitude of $ns^A_{ij}$ and $ns^D_{ij}$ (i.e., number of common sellers $b_i$ and $b_j$ agree or disagree on, respectively) for constructing ${\bf P}_B$ and similarly for ${\bf P}_S$, although we fixed $\delta_p$ and $\delta_n$ for constructing both one-mode projection matrices. We observe in terms of both AUC and F1 from Figure~\ref{fig:SBRW_deltas} that indeed using these two thresholds to avoid implicit links that do not have a significant amount of information (i.e., low magnitude of |$ns^A_{ij}$ - $ns^D_{ij}$|) provides great improvement to our method. It appears that implicit positive links that have low support are helpful to include. However, it seems better to avoid inferred negative links, which we obtained based on balance theory in the form of signed triangles between two buyers and a seller (similarly for the case of two sellers and a buyer). Although including a few is helpful, ones that have low amount of balance theory support from the network are better left out of the propagation process.

\vspace{-1ex}
\section{Related Work}\label{sec:related_work}
Our work is mostly related to signed network analysis~\cite{leskovec2010signed,kunegis2009slashdot,tang2016survey}, which has primarily been studied in the unipartite setting, and also to unsigned bipartite methodologies~\cite{aksoy2017measuring,zweig2011systematic,kunegis2015exploiting}. 

There have been numerous works that focus on sign prediction under the unipartite setting~\cite{chiang2011exploiting, javari2014cluster, zhang2013characterization,papaoikonomou2014edge,leskovec2010predicting,derr2018opinions}. In~\cite{chiang2011exploiting} a supervised classifier was presented exploiting balance theory through cycles of length 3 and greater to predict signs. However, unlike our work, they were focused in the directed unipartite signed network setting. In comparison, we perform a comprehensive analysis where we first extend balance theory to signed bipartite networks and then use our proposed signed butterflies for sign prediction in the setting of having two types of nodes where links are only between differing types. Signed network embedding ~\cite{Wang-etal2017,yuan2017sne,derr2018sgcn,wang2018shine} is a related area of research that seeks to learn a representation for each node in a signed network that can then later be used for a plethora of tasks including sign prediction, node classification, and visualization. However, they are not specifically designed for bipartite networks and those that utilize balance theory primarily harnessed the social theory by leveraging triangles, which would not be applicable here. Another related domain is that of applying signed network analysis to political networks, such as congressional vote analysis~\cite{derr2019congress,derr2018congress,levorato2016brazilian}.   

Although here we perform sign prediction, our work is closely related to link prediction in unsigned bipartite networks~\cite{kunegis2010link,kunegis2015exploiting,li2013recommendation,allali2013internal}. One such approach is a local method similar to common-neighbors, but given for bipartite networks in~\cite{daminelli2015common}. In~\cite{kunegis2010link} numerous methods are provided for predicting links in bipartite networks, such as graph kernels (e.g., Exponential kernel~\cite{Kondor02diffusionkernels}) and a preferential attachment based approach to predict missing links. In~\cite{gao2017projection} they use the one-mode projection matrix to construct a candidate set and only predict links if nodes are among the discovered set of similar nodes. Most recently, a bipartite network embedding algorithm was presented in~\cite{Gao2018bine} that achieves state-of-the-art performance on predicting links in unsigned bipartite networks, by using random walks to extract implicit links to boost performance beyond only using explicit links, which is somewhat similar to the ideas of MFwBT.

\section{Conclusion}\label{sec:conclusion}
In conclusion, signed bipartite networks are a specific type of networks that have become increasingly ubiquitous, but yet fall between the cracks since their added complexities coming from both the negative links and bipartite setting have left both methods and theories lacking the capability to correctly handle them. Meanwhile, balance theory, a key signed social theory, has shown to provide vast improvements in modeling, measuring, and mining tasks related to signed networks when utilized in the form of triangles. Thus, we provided an initial investigation of balance theory in signed bipartite networks that: (1) extends the definition in the form of signed butterfly isomorphism classes; (2) validated that indeed balanced signed butterflies are found significantly more often as compared to unbalanced in signed bipartite networks; (3) leveraged balance theory for the construction of multiple sign prediction methods; and (4) performed experiments on three real-world signed bipartite networks to provide insight into both balance theory and sign prediction in signed bipartite networks.

Our future work will first consist of gaining an even better understanding of the dynamics of signed bipartite networks and how social theories such as balance theory affect their construction/evolution. Thereafter, we plan to utilize signed butterflies for other network analysis tasks in the signed bipartite setting such as network embedding~\cite{derr2018sgcn,Wang-etal2017} and tie strength prediction~\cite{xiang2010modeling}. 
We also plan to further pursue the usefulness of the signed bipartite network formulation of the US Congress for in-depth analysis and prediction tasks such as ``swing votes'', which are votes coming from a representatives that is voting against what their party suggests.

\begin{acks}
Tyler Derr and Jiliang Tang are supported by the National ScienceFoundation (NSF) under grant numbers IIS-1714741, IIS-1715940, IIS-1845081 and CNS-1815636, and a grant from Criteo Faculty Research Award. 
\end{acks}

\balance
\bibliographystyle{ACM-Reference-Format}
\bibliography{references/signed_bipartite,references/sgcn,references/signed,references/signed_network_modeling}


\begin{thebibliography}{51}


\ifx \showCODEN    \undefined \def \showCODEN     #1{\unskip}     \fi
\ifx \showDOI      \undefined \def \showDOI       #1{#1}\fi
\ifx \showISBNx    \undefined \def \showISBNx     #1{\unskip}     \fi
\ifx \showISBNxiii \undefined \def \showISBNxiii  #1{\unskip}     \fi
\ifx \showISSN     \undefined \def \showISSN      #1{\unskip}     \fi
\ifx \showLCCN     \undefined \def \showLCCN      #1{\unskip}     \fi
\ifx \shownote     \undefined \def \shownote      #1{#1}          \fi
\ifx \showarticletitle \undefined \def \showarticletitle #1{#1}   \fi
\ifx \showURL      \undefined \def \showURL       {\relax}        \fi
\providecommand\bibfield[2]{#2}
\providecommand\bibinfo[2]{#2}
\providecommand\natexlab[1]{#1}
\providecommand\showeprint[2][]{arXiv:#2}

\bibitem[\protect\citeauthoryear{Aksoy, Kolda, and Pinar}{Aksoy
  et~al\mbox{.}}{2017}]%
        {aksoy2017measuring}
\bibfield{author}{\bibinfo{person}{Sinan~G Aksoy}, \bibinfo{person}{Tamara~G
  Kolda}, {and} \bibinfo{person}{Ali Pinar}.} \bibinfo{year}{2017}\natexlab{}.
\newblock \showarticletitle{Measuring and modeling bipartite graphs with
  community structure}.
\newblock \bibinfo{journal}{\emph{Journal of Complex Networks}}
  \bibinfo{volume}{5}, \bibinfo{number}{4} (\bibinfo{year}{2017}),
  \bibinfo{pages}{581--603}.
\newblock


\bibitem[\protect\citeauthoryear{Al~Hasan, Chaoji, Salem, and Zaki}{Al~Hasan
  et~al\mbox{.}}{2006}]%
        {al2006link}
\bibfield{author}{\bibinfo{person}{Mohammad Al~Hasan}, \bibinfo{person}{Vineet
  Chaoji}, \bibinfo{person}{Saeed Salem}, {and} \bibinfo{person}{Mohammed
  Zaki}.} \bibinfo{year}{2006}\natexlab{}.
\newblock \showarticletitle{Link prediction using supervised learning}. In
  \bibinfo{booktitle}{\emph{SDM06: workshop on link analysis, counter-terrorism
  and security}}.
\newblock


\bibitem[\protect\citeauthoryear{Allali, Magnien, and Latapy}{Allali
  et~al\mbox{.}}{2013}]%
        {allali2013internal}
\bibfield{author}{\bibinfo{person}{Oussama Allali},
  \bibinfo{person}{Cl{\'e}mence Magnien}, {and} \bibinfo{person}{Matthieu
  Latapy}.} \bibinfo{year}{2013}\natexlab{}.
\newblock \showarticletitle{Internal link prediction: A new approach for
  predicting links in bipartite graphs}.
\newblock \bibinfo{journal}{\emph{Intelligent Data Analysis}}
  \bibinfo{volume}{17}, \bibinfo{number}{1} (\bibinfo{year}{2013}),
  \bibinfo{pages}{5--25}.
\newblock


\bibitem[\protect\citeauthoryear{Anchuri and Magdon-Ismail}{Anchuri and
  Magdon-Ismail}{2012}]%
        {anchuri2012communities}
\bibfield{author}{\bibinfo{person}{Pranay Anchuri} {and} \bibinfo{person}{Malik
  Magdon-Ismail}.} \bibinfo{year}{2012}\natexlab{}.
\newblock \showarticletitle{Communities and balance in signed networks: A
  spectral approach}. In \bibinfo{booktitle}{\emph{Advances in Social Networks
  Analysis and Mining (ASONAM), 2012 IEEE/ACM International Conference on}}.
  IEEE, \bibinfo{pages}{235--242}.
\newblock


\bibitem[\protect\citeauthoryear{Bonacich and Lloyd}{Bonacich and
  Lloyd}{2004}]%
        {bonacich2004calculating}
\bibfield{author}{\bibinfo{person}{Phillip Bonacich} {and}
  \bibinfo{person}{Paulette Lloyd}.} \bibinfo{year}{2004}\natexlab{}.
\newblock \showarticletitle{Calculating status with negative relations}.
\newblock \bibinfo{journal}{\emph{Social Networks}} \bibinfo{volume}{26},
  \bibinfo{number}{4} (\bibinfo{year}{2004}), \bibinfo{pages}{331--338}.
\newblock


\bibitem[\protect\citeauthoryear{Cartwright and Harary}{Cartwright and
  Harary}{1956}]%
        {Cartwright-Haray1956}
\bibfield{author}{\bibinfo{person}{Dorwin Cartwright} {and}
  \bibinfo{person}{Frank Harary}.} \bibinfo{year}{1956}\natexlab{}.
\newblock \showarticletitle{Structural balance: a generalization of Heider's
  theory.}
\newblock \bibinfo{journal}{\emph{Psychological review}} \bibinfo{volume}{63},
  \bibinfo{number}{5} (\bibinfo{year}{1956}), \bibinfo{pages}{277}.
\newblock


\bibitem[\protect\citeauthoryear{Chiang, Natarajan, Tewari, and Dhillon}{Chiang
  et~al\mbox{.}}{2011}]%
        {chiang2011exploiting}
\bibfield{author}{\bibinfo{person}{Kai-Yang Chiang}, \bibinfo{person}{Nagarajan
  Natarajan}, \bibinfo{person}{Ambuj Tewari}, {and} \bibinfo{person}{Inderjit~S
  Dhillon}.} \bibinfo{year}{2011}\natexlab{}.
\newblock \showarticletitle{Exploiting longer cycles for link prediction in
  signed networks}. In \bibinfo{booktitle}{\emph{Proceedings of the 20th ACM
  international conference on Information and knowledge management}}. ACM,
  \bibinfo{pages}{1157--1162}.
\newblock


\bibitem[\protect\citeauthoryear{Daminelli, Thomas, Dur{\'a}n, and
  Cannistraci}{Daminelli et~al\mbox{.}}{2015}]%
        {daminelli2015common}
\bibfield{author}{\bibinfo{person}{Simone Daminelli},
  \bibinfo{person}{Josephine~Maria Thomas}, \bibinfo{person}{Claudio
  Dur{\'a}n}, {and} \bibinfo{person}{Carlo~Vittorio Cannistraci}.}
  \bibinfo{year}{2015}\natexlab{}.
\newblock \showarticletitle{Common neighbours and the local-community-paradigm
  for topological link prediction in bipartite networks}.
\newblock \bibinfo{journal}{\emph{New Journal of Physics}}
  \bibinfo{volume}{17}, \bibinfo{number}{11} (\bibinfo{year}{2015}),
  \bibinfo{pages}{113037}.
\newblock


\bibitem[\protect\citeauthoryear{Derr, Aggarwal, and Tang}{Derr
  et~al\mbox{.}}{2018a}]%
        {derr2018signed}
\bibfield{author}{\bibinfo{person}{Tyler Derr}, \bibinfo{person}{Charu
  Aggarwal}, {and} \bibinfo{person}{Jiliang Tang}.}
  \bibinfo{year}{2018}\natexlab{a}.
\newblock \showarticletitle{Signed Network Modeling Based on Structural Balance
  Theory}. In \bibinfo{booktitle}{\emph{Proceedings of the 27th ACM
  International Conference on Information and Knowledge Management}}. ACM,
  \bibinfo{pages}{557--566}.
\newblock


\bibitem[\protect\citeauthoryear{Derr, Ma, and Tang}{Derr
  et~al\mbox{.}}{2018b}]%
        {derr2018sgcn}
\bibfield{author}{\bibinfo{person}{Tyler Derr}, \bibinfo{person}{Yao Ma}, {and}
  \bibinfo{person}{Jiliang Tang}.} \bibinfo{year}{2018}\natexlab{b}.
\newblock \showarticletitle{Signed graph convolutional networks}. In
  \bibinfo{booktitle}{\emph{2018 IEEE International Conference on Data Mining
  (ICDM)}}. IEEE, \bibinfo{pages}{929--934}.
\newblock


\bibitem[\protect\citeauthoryear{Derr and Tang}{Derr and Tang}{2018}]%
        {derr2018congress}
\bibfield{author}{\bibinfo{person}{Tyler Derr} {and} \bibinfo{person}{Jiliang
  Tang}.} \bibinfo{year}{2018}\natexlab{}.
\newblock \showarticletitle{Congressional Vote Analysis Using Signed Networks}.
  In \bibinfo{booktitle}{\emph{2018 IEEE International Conference on Data
  Mining Workshops (ICDMW)}}. IEEE, \bibinfo{pages}{1501--1502}.
\newblock


\bibitem[\protect\citeauthoryear{Derr, Wang, Wang, and Tang}{Derr
  et~al\mbox{.}}{2018d}]%
        {derr2018relevance}
\bibfield{author}{\bibinfo{person}{Tyler Derr}, \bibinfo{person}{Chenxing
  Wang}, \bibinfo{person}{Suhang Wang}, {and} \bibinfo{person}{Jiliang Tang}.}
  \bibinfo{year}{2018}\natexlab{d}.
\newblock \showarticletitle{Relevance Measurements in Online Signed Social
  Networks}. In \bibinfo{booktitle}{\emph{Proceedings of the 14th International
  Workshop on Mining and Learning with Graphs (MLG)}}.
\newblock


\bibitem[\protect\citeauthoryear{Derr, Wang, and Tang}{Derr
  et~al\mbox{.}}{2018c}]%
        {derr2018opinions}
\bibfield{author}{\bibinfo{person}{Tyler Derr}, \bibinfo{person}{Zhiwei Wang},
  {and} \bibinfo{person}{Jiliang Tang}.} \bibinfo{year}{2018}\natexlab{c}.
\newblock \showarticletitle{Opinions Power Opinions: Joint Link and Interaction
  Polarity Prediction in Signed Networks}. In
  \bibinfo{booktitle}{\emph{Advances in Social Networks Analysis and Mining
  (ASONAM), 2018 IEEE/ACM International Conference on}}. IEEE.
\newblock


\bibitem[\protect\citeauthoryear{Forsati, Barjasteh, Masrour, Esfahanian, and
  Radha}{Forsati et~al\mbox{.}}{2015}]%
        {forsati2015pushtrust}
\bibfield{author}{\bibinfo{person}{Rana Forsati}, \bibinfo{person}{Iman
  Barjasteh}, \bibinfo{person}{Farzan Masrour}, \bibinfo{person}{Abdol-Hossein
  Esfahanian}, {and} \bibinfo{person}{Hayder Radha}.}
  \bibinfo{year}{2015}\natexlab{}.
\newblock \showarticletitle{Pushtrust: An efficient recommendation algorithm by
  leveraging trust and distrust relations}. In
  \bibinfo{booktitle}{\emph{Proceedings of the 9th ACM Conference on
  Recommender Systems}}. ACM, \bibinfo{pages}{51--58}.
\newblock


\bibitem[\protect\citeauthoryear{Gao, Chen, He, and Zhou}{Gao
  et~al\mbox{.}}{2018}]%
        {Gao2018bine}
\bibfield{author}{\bibinfo{person}{Ming Gao}, \bibinfo{person}{Leihui Chen},
  \bibinfo{person}{Xiangnan He}, {and} \bibinfo{person}{Aoying Zhou}.}
  \bibinfo{year}{2018}\natexlab{}.
\newblock \showarticletitle{BiNE: Bipartite Network Embedding}. In
  \bibinfo{booktitle}{\emph{The 41st International ACM SIGIR Conference on
  Research \&\#38; Development in Information Retrieval}}
  \emph{(\bibinfo{series}{SIGIR '18})}. \bibinfo{publisher}{ACM},
  \bibinfo{address}{New York, NY, USA}, \bibinfo{pages}{715--724}.
\newblock


\bibitem[\protect\citeauthoryear{Gao, Chen, Li, Li, Liu, and Xu}{Gao
  et~al\mbox{.}}{2017}]%
        {gao2017projection}
\bibfield{author}{\bibinfo{person}{Man Gao}, \bibinfo{person}{Ling Chen},
  \bibinfo{person}{Bin Li}, \bibinfo{person}{Yun Li}, \bibinfo{person}{Wei
  Liu}, {and} \bibinfo{person}{Yong-cheng Xu}.}
  \bibinfo{year}{2017}\natexlab{}.
\newblock \showarticletitle{Projection-based link prediction in a bipartite
  network}.
\newblock \bibinfo{journal}{\emph{Information Sciences}}  \bibinfo{volume}{376}
  (\bibinfo{year}{2017}), \bibinfo{pages}{158--171}.
\newblock


\bibitem[\protect\citeauthoryear{Heider}{Heider}{1946}]%
        {heider1946attitudes}
\bibfield{author}{\bibinfo{person}{Fritz Heider}.}
  \bibinfo{year}{1946}\natexlab{}.
\newblock \showarticletitle{Attitudes and cognitive organization}.
\newblock \bibinfo{journal}{\emph{The Journal of psychology}}
  \bibinfo{volume}{21}, \bibinfo{number}{1} (\bibinfo{year}{1946}),
  \bibinfo{pages}{107--112}.
\newblock


\bibitem[\protect\citeauthoryear{Hsieh, Chiang, and Dhillon}{Hsieh
  et~al\mbox{.}}{2012}]%
        {hsieh2012low}
\bibfield{author}{\bibinfo{person}{Cho-Jui Hsieh}, \bibinfo{person}{Kai-Yang
  Chiang}, {and} \bibinfo{person}{Inderjit~S Dhillon}.}
  \bibinfo{year}{2012}\natexlab{}.
\newblock \showarticletitle{Low rank modeling of signed networks}. In
  \bibinfo{booktitle}{\emph{Proceedings of the 18th ACM SIGKDD international
  conference on Knowledge discovery and data mining}}. ACM,
  \bibinfo{pages}{507--515}.
\newblock


\bibitem[\protect\citeauthoryear{Javari and Jalili}{Javari and Jalili}{2014}]%
        {javari2014cluster}
\bibfield{author}{\bibinfo{person}{Amin Javari} {and} \bibinfo{person}{Mahdi
  Jalili}.} \bibinfo{year}{2014}\natexlab{}.
\newblock \showarticletitle{Cluster-Based Collaborative Filtering for Sign
  Prediction in Social Networks with Positive and Negative Links}.
\newblock \bibinfo{journal}{\emph{ACM Transactions on Intelligent Systems and
  Technology (TIST)}} \bibinfo{volume}{5}, \bibinfo{number}{2}
  (\bibinfo{year}{2014}), \bibinfo{pages}{24}.
\newblock


\bibitem[\protect\citeauthoryear{Jung, Jin, Sael, and Kang}{Jung
  et~al\mbox{.}}{2016}]%
        {Jung2016srwr}
\bibfield{author}{\bibinfo{person}{Jinhong Jung}, \bibinfo{person}{Woojeong
  Jin}, \bibinfo{person}{Lee Sael}, {and} \bibinfo{person}{U. Kang}.}
  \bibinfo{year}{2016}\natexlab{}.
\newblock \showarticletitle{Personalized Ranking in Signed Networks Using
  Signed Random Walk with Restart}. In \bibinfo{booktitle}{\emph{{IEEE} 16th
  International Conference on Data Mining, {ICDM} 2016, December 12-15, 2016,
  Barcelona, Spain}}. \bibinfo{pages}{973--978}.
\newblock
\urldef\tempurl%
\url{https://doi.org/10.1109/ICDM.2016.0122}
\showDOI{\tempurl}


\bibitem[\protect\citeauthoryear{Karimi, Derr, Brookhouse, and Tang}{Karimi
  et~al\mbox{.}}{2019}]%
        {derr2019congress}
\bibfield{author}{\bibinfo{person}{Hamid Karimi}, \bibinfo{person}{Tyler Derr},
  \bibinfo{person}{Aaron Brookhouse}, {and} \bibinfo{person}{Jiliang Tang}.}
  \bibinfo{year}{2019}\natexlab{}.
\newblock \showarticletitle{Multi-Factor Congressional Vote Prediction}. In
  \bibinfo{booktitle}{\emph{Advances in Social Networks Analysis and Mining
  (ASONAM), 2019 IEEE/ACM International Conference on}}. IEEE.
\newblock


\bibitem[\protect\citeauthoryear{Kondor and Lafferty}{Kondor and
  Lafferty}{2002}]%
        {Kondor02diffusionkernels}
\bibfield{author}{\bibinfo{person}{Risi~Imre Kondor} {and}
  \bibinfo{person}{John Lafferty}.} \bibinfo{year}{2002}\natexlab{}.
\newblock \showarticletitle{Diffusion kernels on graphs and other discrete
  input spaces}. In \bibinfo{booktitle}{\emph{Proceedings of the 19th
  International Conference on Machine Learning}}. \bibinfo{publisher}{Morgan
  Kaufmann}, \bibinfo{pages}{315--322}.
\newblock


\bibitem[\protect\citeauthoryear{Kumar, Spezzano, Subrahmanian, and
  Faloutsos}{Kumar et~al\mbox{.}}{2016}]%
        {kumar2016edge}
\bibfield{author}{\bibinfo{person}{Srijan Kumar}, \bibinfo{person}{Francesca
  Spezzano}, \bibinfo{person}{VS Subrahmanian}, {and} \bibinfo{person}{Christos
  Faloutsos}.} \bibinfo{year}{2016}\natexlab{}.
\newblock \showarticletitle{Edge Weight Prediction in Weighted Signed
  Networks.}. In \bibinfo{booktitle}{\emph{ICDM}}. \bibinfo{pages}{221--230}.
\newblock


\bibitem[\protect\citeauthoryear{Kunegis}{Kunegis}{2015}]%
        {kunegis2015exploiting}
\bibfield{author}{\bibinfo{person}{J{\'e}r{\^o}me Kunegis}.}
  \bibinfo{year}{2015}\natexlab{}.
\newblock \showarticletitle{Exploiting the structure of bipartite graphs for
  algebraic and spectral graph theory applications}.
\newblock \bibinfo{journal}{\emph{Internet Mathematics}} \bibinfo{volume}{11},
  \bibinfo{number}{3} (\bibinfo{year}{2015}), \bibinfo{pages}{201--321}.
\newblock


\bibitem[\protect\citeauthoryear{Kunegis, De~Luca, and Albayrak}{Kunegis
  et~al\mbox{.}}{2010}]%
        {kunegis2010link}
\bibfield{author}{\bibinfo{person}{J{\'e}r{\^o}me Kunegis},
  \bibinfo{person}{Ernesto~W De~Luca}, {and} \bibinfo{person}{Sahin Albayrak}.}
  \bibinfo{year}{2010}\natexlab{}.
\newblock \showarticletitle{The link prediction problem in bipartite networks}.
  In \bibinfo{booktitle}{\emph{International Conference on Information
  Processing and Management of Uncertainty in Knowledge-based Systems}}.
  Springer, \bibinfo{pages}{380--389}.
\newblock


\bibitem[\protect\citeauthoryear{Kunegis, Lommatzsch, and Bauckhage}{Kunegis
  et~al\mbox{.}}{2009}]%
        {kunegis2009slashdot}
\bibfield{author}{\bibinfo{person}{J{\'e}r{\^o}me Kunegis},
  \bibinfo{person}{Andreas Lommatzsch}, {and} \bibinfo{person}{Christian
  Bauckhage}.} \bibinfo{year}{2009}\natexlab{}.
\newblock \showarticletitle{The slashdot zoo: mining a social network with
  negative edges}. In \bibinfo{booktitle}{\emph{Proceedings of the 18th
  international conference on World wide web}}. ACM, \bibinfo{pages}{741--750}.
\newblock


\bibitem[\protect\citeauthoryear{Leskovec, Huttenlocher, and
  Kleinberg}{Leskovec et~al\mbox{.}}{2010a}]%
        {leskovec2010predicting}
\bibfield{author}{\bibinfo{person}{Jure Leskovec}, \bibinfo{person}{Daniel
  Huttenlocher}, {and} \bibinfo{person}{Jon Kleinberg}.}
  \bibinfo{year}{2010}\natexlab{a}.
\newblock \showarticletitle{Predicting positive and negative links in online
  social networks}. In \bibinfo{booktitle}{\emph{Proceedings of the 19th
  international conference on World wide web}}. ACM, \bibinfo{pages}{641--650}.
\newblock


\bibitem[\protect\citeauthoryear{Leskovec, Huttenlocher, and
  Kleinberg}{Leskovec et~al\mbox{.}}{2010b}]%
        {leskovec2010signed}
\bibfield{author}{\bibinfo{person}{Jure Leskovec}, \bibinfo{person}{Daniel
  Huttenlocher}, {and} \bibinfo{person}{Jon Kleinberg}.}
  \bibinfo{year}{2010}\natexlab{b}.
\newblock \showarticletitle{Signed networks in social media}. In
  \bibinfo{booktitle}{\emph{Proceedings of the SIGCHI Conference on Human
  Factors in Computing Systems}}. ACM, \bibinfo{pages}{1361--1370}.
\newblock


\bibitem[\protect\citeauthoryear{Levorato and Frota}{Levorato and
  Frota}{2016}]%
        {levorato2016brazilian}
\bibfield{author}{\bibinfo{person}{Mario Levorato} {and} \bibinfo{person}{Yuri
  Frota}.} \bibinfo{year}{2016}\natexlab{}.
\newblock \showarticletitle{Brazilian Congress structural balance analysis}.
\newblock \bibinfo{journal}{\emph{arXiv preprint arXiv:1609.00767}}
  (\bibinfo{year}{2016}).
\newblock


\bibitem[\protect\citeauthoryear{Li and Chen}{Li and Chen}{2013}]%
        {li2013recommendation}
\bibfield{author}{\bibinfo{person}{Xin Li} {and} \bibinfo{person}{Hsinchun
  Chen}.} \bibinfo{year}{2013}\natexlab{}.
\newblock \showarticletitle{Recommendation as link prediction in bipartite
  graphs: A graph kernel-based machine learning approach}.
\newblock \bibinfo{journal}{\emph{Decision Support Systems}}
  \bibinfo{volume}{54}, \bibinfo{number}{2} (\bibinfo{year}{2013}),
  \bibinfo{pages}{880--890}.
\newblock


\bibitem[\protect\citeauthoryear{Lov{\'a}sz et~al\mbox{.}}{Lov{\'a}sz
  et~al\mbox{.}}{1993}]%
        {lovasz1993random}
\bibfield{author}{\bibinfo{person}{L{\'a}szl{\'o} Lov{\'a}sz} {et~al\mbox{.}}}
  \bibinfo{year}{1993}\natexlab{}.
\newblock \showarticletitle{Random walks on graphs: A survey}.
\newblock \bibinfo{journal}{\emph{Combinatorics, Paul erdos is eighty}}
  \bibinfo{volume}{2}, \bibinfo{number}{1} (\bibinfo{year}{1993}),
  \bibinfo{pages}{1--46}.
\newblock


\bibitem[\protect\citeauthoryear{Lu, Savas, Tang, and Dhillon}{Lu
  et~al\mbox{.}}{2010}]%
        {lu2010supervised}
\bibfield{author}{\bibinfo{person}{Zhengdong Lu}, \bibinfo{person}{Berkant
  Savas}, \bibinfo{person}{Wei Tang}, {and} \bibinfo{person}{Inderjit~S
  Dhillon}.} \bibinfo{year}{2010}\natexlab{}.
\newblock \showarticletitle{Supervised link prediction using multiple sources}.
  In \bibinfo{booktitle}{\emph{Data Mining (ICDM), 2010 IEEE 10th International
  Conference on}}. IEEE, \bibinfo{pages}{923--928}.
\newblock


\bibitem[\protect\citeauthoryear{Ludwig and Abell}{Ludwig and Abell}{2007}]%
        {ludwig2007evolutionary}
\bibfield{author}{\bibinfo{person}{Mark Ludwig} {and} \bibinfo{person}{Peter
  Abell}.} \bibinfo{year}{2007}\natexlab{}.
\newblock \showarticletitle{An evolutionary model of social networks.}
\newblock \bibinfo{journal}{\emph{European Physical Journal B--Condensed
  Matter}} \bibinfo{volume}{58}, \bibinfo{number}{1} (\bibinfo{year}{2007}).
\newblock


\bibitem[\protect\citeauthoryear{Menon and Elkan}{Menon and Elkan}{2011}]%
        {menon2011link}
\bibfield{author}{\bibinfo{person}{Aditya~Krishna Menon} {and}
  \bibinfo{person}{Charles Elkan}.} \bibinfo{year}{2011}\natexlab{}.
\newblock \showarticletitle{Link prediction via matrix factorization}. In
  \bibinfo{booktitle}{\emph{Joint european conference on machine learning and
  knowledge discovery in databases}}. Springer, \bibinfo{pages}{437--452}.
\newblock


\bibitem[\protect\citeauthoryear{Papaoikonomou, Kardara, Tserpes, and
  Varvarigou}{Papaoikonomou et~al\mbox{.}}{2014}]%
        {papaoikonomou2014edge}
\bibfield{author}{\bibinfo{person}{Athanasios Papaoikonomou},
  \bibinfo{person}{Magdalini Kardara}, \bibinfo{person}{Konstantinos Tserpes},
  {and} \bibinfo{person}{Dora Varvarigou}.} \bibinfo{year}{2014}\natexlab{}.
\newblock \showarticletitle{Edge Sign Prediction in Social Networks via
  Frequent Subgraph Discovery}.
\newblock \bibinfo{journal}{\emph{IEEE Internet Computing}}
  (\bibinfo{year}{2014}).
\newblock


\bibitem[\protect\citeauthoryear{Sanei-Mehri, Sariyuce, and
  Tirthapura}{Sanei-Mehri et~al\mbox{.}}{2018}]%
        {sanei2018butterfly}
\bibfield{author}{\bibinfo{person}{Seyed-Vahid Sanei-Mehri},
  \bibinfo{person}{Ahmet~Erdem Sariyuce}, {and} \bibinfo{person}{Srikanta
  Tirthapura}.} \bibinfo{year}{2018}\natexlab{}.
\newblock \showarticletitle{Butterfly Counting in Bipartite Networks}. In
  \bibinfo{booktitle}{\emph{Proceedings of the 24th ACM SIGKDD International
  Conference on Knowledge Discovery \& Data Mining}}. ACM,
  \bibinfo{pages}{2150--2159}.
\newblock


\bibitem[\protect\citeauthoryear{Shahriari and Jalili}{Shahriari and
  Jalili}{2014}]%
        {Shahriari-Jalili2014}
\bibfield{author}{\bibinfo{person}{Moshen Shahriari} {and}
  \bibinfo{person}{Mahdi Jalili}.} \bibinfo{year}{2014}\natexlab{}.
\newblock \showarticletitle{Ranking nodes in signed social networks}.
\newblock \bibinfo{journal}{\emph{Social Network Analysis and Mining}}
  \bibinfo{volume}{4}, \bibinfo{number}{1} (\bibinfo{year}{2014}),
  \bibinfo{pages}{172}.
\newblock


\bibitem[\protect\citeauthoryear{Tang, Aggarwal, and Liu}{Tang
  et~al\mbox{.}}{2016a}]%
        {tang2016recommendations}
\bibfield{author}{\bibinfo{person}{Jiliang Tang}, \bibinfo{person}{Charu
  Aggarwal}, {and} \bibinfo{person}{Huan Liu}.}
  \bibinfo{year}{2016}\natexlab{a}.
\newblock \showarticletitle{Recommendations in signed social networks}. In
  \bibinfo{booktitle}{\emph{Proceedings of the 25th International Conference on
  World Wide Web}}. International World Wide Web Conferences Steering
  Committee, \bibinfo{pages}{31--40}.
\newblock


\bibitem[\protect\citeauthoryear{Tang, Chang, Aggarwal, and Liu}{Tang
  et~al\mbox{.}}{2016b}]%
        {tang2016survey}
\bibfield{author}{\bibinfo{person}{Jiliang Tang}, \bibinfo{person}{Yi Chang},
  \bibinfo{person}{Charu Aggarwal}, {and} \bibinfo{person}{Huan Liu}.}
  \bibinfo{year}{2016}\natexlab{b}.
\newblock \showarticletitle{A survey of signed network mining in social media}.
\newblock \bibinfo{journal}{\emph{ACM Computing Surveys (CSUR)}}
  \bibinfo{volume}{49}, \bibinfo{number}{3} (\bibinfo{year}{2016}),
  \bibinfo{pages}{42}.
\newblock


\bibitem[\protect\citeauthoryear{Tong, Faloutsos, and Pan}{Tong
  et~al\mbox{.}}{2006}]%
        {tong2006fast}
\bibfield{author}{\bibinfo{person}{Hanghang Tong}, \bibinfo{person}{Christos
  Faloutsos}, {and} \bibinfo{person}{Jia-yu Pan}.}
  \bibinfo{year}{2006}\natexlab{}.
\newblock \showarticletitle{Fast Random Walk with Restart and Its
  Applications}. In \bibinfo{booktitle}{\emph{Data Mining, 2006. ICDM'06. Sixth
  International Conference on}}. IEEE, \bibinfo{pages}{613--622}.
\newblock


\bibitem[\protect\citeauthoryear{Traag, Nesterov, and Van~Dooren}{Traag
  et~al\mbox{.}}{2010}]%
        {traag2010exponential}
\bibfield{author}{\bibinfo{person}{Vincent Traag}, \bibinfo{person}{Yurii
  Nesterov}, {and} \bibinfo{person}{Paul Van~Dooren}.}
  \bibinfo{year}{2010}\natexlab{}.
\newblock \showarticletitle{Exponential Ranking: Taking into Account Negative
  Links}.
\newblock \bibinfo{journal}{\emph{Social Informatics}} (\bibinfo{year}{2010}),
  \bibinfo{pages}{192--202}.
\newblock


\bibitem[\protect\citeauthoryear{Vuka{\v{s}}inovi{\'c}, {\v{S}}ilc, and
  {\v{S}}krekovski}{Vuka{\v{s}}inovi{\'c} et~al\mbox{.}}{2014}]%
        {vukavsinovic2014modeling}
\bibfield{author}{\bibinfo{person}{Vida Vuka{\v{s}}inovi{\'c}},
  \bibinfo{person}{Jurij {\v{S}}ilc}, {and} \bibinfo{person}{Risth
  {\v{S}}krekovski}.} \bibinfo{year}{2014}\natexlab{}.
\newblock \showarticletitle{Modeling acquaintance networks based on balance
  theory}.
\newblock \bibinfo{journal}{\emph{International Journal of Applied Mathematics
  and Computer Science}} \bibinfo{volume}{24}, \bibinfo{number}{3}
  (\bibinfo{year}{2014}), \bibinfo{pages}{683--696}.
\newblock


\bibitem[\protect\citeauthoryear{Wang, Zhang, Hou, Xie, Guo, and Liu}{Wang
  et~al\mbox{.}}{2018}]%
        {wang2018shine}
\bibfield{author}{\bibinfo{person}{Hongwei Wang}, \bibinfo{person}{Fuzheng
  Zhang}, \bibinfo{person}{Min Hou}, \bibinfo{person}{Xing Xie},
  \bibinfo{person}{Minyi Guo}, {and} \bibinfo{person}{Qi Liu}.}
  \bibinfo{year}{2018}\natexlab{}.
\newblock \showarticletitle{SHINE: signed heterogeneous information network
  embedding for sentiment link prediction}. In
  \bibinfo{booktitle}{\emph{Proceedings of the Eleventh ACM International
  Conference on Web Search and Data Mining}}. ACM, \bibinfo{pages}{592--600}.
\newblock


\bibitem[\protect\citeauthoryear{Wang, Tang, Aggarwal, Chang, and Liu}{Wang
  et~al\mbox{.}}{2017}]%
        {Wang-etal2017}
\bibfield{author}{\bibinfo{person}{Suhang Wang}, \bibinfo{person}{Jiliang
  Tang}, \bibinfo{person}{Charu Aggarwal}, \bibinfo{person}{Yi Chang}, {and}
  \bibinfo{person}{Huan Liu}.} \bibinfo{year}{2017}\natexlab{}.
\newblock \showarticletitle{Signed network embedding in social media}. In
  \bibinfo{booktitle}{\emph{Proceedings of the 2017 SIAM International
  Conference on Data Mining}}. SIAM, \bibinfo{pages}{327--335}.
\newblock


\bibitem[\protect\citeauthoryear{Wu, Aggarwal, and Sun}{Wu
  et~al\mbox{.}}{2016}]%
        {Wu-etal2016}
\bibfield{author}{\bibinfo{person}{Zhaoming Wu}, \bibinfo{person}{Charu~C
  Aggarwal}, {and} \bibinfo{person}{Jimeng Sun}.}
  \bibinfo{year}{2016}\natexlab{}.
\newblock \showarticletitle{The troll-trust model for ranking in signed
  networks}. In \bibinfo{booktitle}{\emph{Proceedings of the Ninth ACM
  International Conference on Web Search and Data Mining}}. ACM,
  \bibinfo{pages}{447--456}.
\newblock


\bibitem[\protect\citeauthoryear{Xiang, Neville, and Rogati}{Xiang
  et~al\mbox{.}}{2010}]%
        {xiang2010modeling}
\bibfield{author}{\bibinfo{person}{Rongjing Xiang}, \bibinfo{person}{Jennifer
  Neville}, {and} \bibinfo{person}{Monica Rogati}.}
  \bibinfo{year}{2010}\natexlab{}.
\newblock \showarticletitle{Modeling relationship strength in online social
  networks}. In \bibinfo{booktitle}{\emph{Proceedings of the 19th international
  conference on World wide web}}. ACM, \bibinfo{pages}{981--990}.
\newblock


\bibitem[\protect\citeauthoryear{Yildirim and Coscia}{Yildirim and
  Coscia}{2014}]%
        {yildirim2014using}
\bibfield{author}{\bibinfo{person}{Muhammed~A Yildirim} {and}
  \bibinfo{person}{Michele Coscia}.} \bibinfo{year}{2014}\natexlab{}.
\newblock \showarticletitle{Using random walks to generate associations between
  objects}.
\newblock \bibinfo{journal}{\emph{PloS one}} \bibinfo{volume}{9},
  \bibinfo{number}{8} (\bibinfo{year}{2014}), \bibinfo{pages}{e104813}.
\newblock


\bibitem[\protect\citeauthoryear{Yuan, Wu, and Xiang}{Yuan
  et~al\mbox{.}}{2017}]%
        {yuan2017sne}
\bibfield{author}{\bibinfo{person}{Shuhan Yuan}, \bibinfo{person}{Xintao Wu},
  {and} \bibinfo{person}{Yang Xiang}.} \bibinfo{year}{2017}\natexlab{}.
\newblock \showarticletitle{SNE: signed network embedding}. In
  \bibinfo{booktitle}{\emph{Pacific-Asia conference on knowledge discovery and
  data mining}}. Springer, \bibinfo{pages}{183--195}.
\newblock


\bibitem[\protect\citeauthoryear{Zhang, Jiang, Bao, and Zhang}{Zhang
  et~al\mbox{.}}{2013}]%
        {zhang2013characterization}
\bibfield{author}{\bibinfo{person}{Tongda Zhang}, \bibinfo{person}{Haomiao
  Jiang}, \bibinfo{person}{Zhouxiao Bao}, {and} \bibinfo{person}{Yingfeng
  Zhang}.} \bibinfo{year}{2013}\natexlab{}.
\newblock \showarticletitle{Characterization and edge sign prediction in signed
  networks}.
\newblock \bibinfo{journal}{\emph{Journal of Industrial and Intelligent
  Information Vol}} \bibinfo{volume}{1}, \bibinfo{number}{1}
  (\bibinfo{year}{2013}).
\newblock


\bibitem[\protect\citeauthoryear{Zhou, Ren, Medo, and Zhang}{Zhou
  et~al\mbox{.}}{2007}]%
        {zhou2007bipartite}
\bibfield{author}{\bibinfo{person}{Tao Zhou}, \bibinfo{person}{Jie Ren},
  \bibinfo{person}{Mat{\'u}{\v{s}} Medo}, {and} \bibinfo{person}{Yi-Cheng
  Zhang}.} \bibinfo{year}{2007}\natexlab{}.
\newblock \showarticletitle{Bipartite network projection and personal
  recommendation}.
\newblock \bibinfo{journal}{\emph{Physical Review E}} \bibinfo{volume}{76},
  \bibinfo{number}{4} (\bibinfo{year}{2007}), \bibinfo{pages}{046115}.
\newblock


\bibitem[\protect\citeauthoryear{Zweig and Kaufmann}{Zweig and
  Kaufmann}{2011}]%
        {zweig2011systematic}
\bibfield{author}{\bibinfo{person}{Katharina~Anna Zweig} {and}
  \bibinfo{person}{Michael Kaufmann}.} \bibinfo{year}{2011}\natexlab{}.
\newblock \showarticletitle{A systematic approach to the one-mode projection of
  bipartite graphs}.
\newblock \bibinfo{journal}{\emph{Social Network Analysis and Mining}}
  \bibinfo{volume}{1}, \bibinfo{number}{3} (\bibinfo{year}{2011}),
  \bibinfo{pages}{187--218}.
\newblock


\end{thebibliography}

\end{document}